\documentclass[aps,rsi,reprint,amsmath,amssymb]{revtex4-2}
\usepackage{graphicx}
\usepackage{dcolumn}
\usepackage{bm}
\usepackage{listings}
\usepackage{color}
\usepackage{multirow}
\usepackage{array}
\usepackage{comment}
\usepackage[utf8]{inputenc}
\usepackage{ifthen}
\usepackage[normalem]{ulem}
\usepackage{amssymb}
\usepackage{amsmath}
\usepackage{bbm}
\usepackage{graphicx}
\usepackage{xcolor}
\usepackage{epstopdf}
\usepackage{multirow}
\usepackage[justification=centering]{caption}
\usepackage{algorithmic,algorithm}
\usepackage[hidelinks]{hyperref}
\usepackage{subfig}
\usepackage[export]{adjustbox}
\usepackage{bm}
\usepackage{enumitem}
\usepackage{amsthm}
\usepackage{epstopdf}

\usepackage{soul}
\usepackage[breakable,skins]{tcolorbox}
\providecommand{\selectlanguage}[1]{}
\renewcommand{\selectlanguage}[1]{}

\newtcolorbox{reviewquote}{enhanced, frame hidden, colback=gray!15, left = 0.5em, right = 0.5em, top = 1ex, bottom = 1ex, borderline west = {2pt} {0pt} {black!20}, breakable}

\definecolor{codegreen}{rgb}{0,0.6,0}
\definecolor{codegray}{rgb}{0.5,0.5,0.5}
\definecolor{codepurple}{rgb}{0.58,0,0.82}
\definecolor{backcolour}{rgb}{0.95,0.95,0.92}

\lstdefinestyle{pythonstyle}{
    backgroundcolor=\color{backcolour},
    commentstyle=\color{codegreen},
    keywordstyle=\color{magenta},
    numberstyle=\tiny\color{codegray},
    stringstyle=\color{codepurple},
    basicstyle=\ttfamily\footnotesize,
    breakatwhitespace=false,
    breaklines=true,
    captionpos=b,
    keepspaces=true,
    numbers=left,
    numbersep=5pt,
    showspaces=false,
    showstringspaces=false,
    showtabs=false,
    tabsize=2,
    language=Python
}

\newboolean{revising}
\setboolean{revising}{true} 

\ifthenelse{\boolean{revising}}
{
    \newcommand{\deleted}[1]{\textcolor{red!80!black}{\st{#1}}} 
} {

    \newcommand{\deleted}[1]{}
}

\begin{document}

\title{SCULPT: An Interactive Machine Learning Platform for Analyzing Multi-Particle Coincidence Data from Cold Target Recoil Ion Momentum Spectroscopy}

\author{Hazem Daoud}
\email{HDaoud@lbl.gov}
\affiliation{Chemical Sciences Division, Atomic, Molecular and Optical Physics, Lawrence Berkeley National Laboratory, 1 Cyclotron Road, Berkeley, California 94720, USA}

\author{Sarvesh Kumar}
\affiliation{Chemical Sciences Division, Atomic, Molecular and Optical Physics, Lawrence Berkeley National Laboratory, 1 Cyclotron Road, Berkeley, California 94720, USA}

\author{Jin Qian}
\affiliation{Chemical Sciences Division, Atomic, Molecular and Optical Physics, Lawrence Berkeley National Laboratory, 1 Cyclotron Road, Berkeley, California 94720, USA}

\author{Tanny Chavez}
\email{TanChavez@lbl.gov}
\affiliation{Advanced Light Source, Lawrence Berkeley National Laboratory, 1 Cyclotron Road, Berkeley, California 94720, USA}

\author{Daniel Slaughter}
\affiliation{Chemical Sciences Division, Atomic, Molecular and Optical Physics, Lawrence Berkeley National Laboratory, 1 Cyclotron Road, Berkeley, California 94720, USA}

\author{Thorsten Weber}
\email{TWeber@lbl.gov}
\affiliation{Chemical Sciences Division, Atomic, Molecular and Optical Physics, Lawrence Berkeley National Laboratory, 1 Cyclotron Road, Berkeley, California 94720, USA}

\date{\today}

\begin{abstract}
We present SCULPT (\textcolor{black}{\textbf{S}}upervised \textcolor{black}{\textbf{C}}lustering and \textcolor{black}{\textbf{U}}ncovering \textcolor{black}{\textbf{L}}atent \textcolor{black}{\textbf{P}}atterns with \textcolor{black}{\textbf{T}}raining), a comprehensive software platform for analyzing tabulated high-dimensional multi-particle coincidence data from Cold Target Recoil Ion Momentum Spectroscopy (COLTRIMS) experiments. The software addresses critical challenges in modern momentum spectroscopy by integrating advanced machine learning techniques with physics-informed analysis in an interactive web-based environment. SCULPT implements Uniform Manifold Approximation and Projection (UMAP) for non-linear dimensionality reduction to reveal correlations in highly dimensional data. We also discuss potential extensions to deep autoencoders for feature learning, and genetic programming for automated discovery of physically meaningful observables. A novel adaptive confidence scoring system provides quantitative reliability 
assessments by evaluating user-selected clustering quality metrics with 
predefined weights that reflect each metric's robustness. The platform features configurable molecular profiles for different experimental systems, interactive visualization with selection tools, and comprehensive data filtering capabilities. Utilizing a subset of SCULPT's capabilities, we analyze photo double ionization data measured using the COLTRIMS method for 3-body dissociation of the D$_2$O molecule, revealing distinct fragmentation channels and their correlations with physics parameters. The software's modular architecture and web-based implementation make it accessible to the broader atomic and molecular physics community, significantly reducing the time required for complex multi-dimensional analyses. This opens the door to finding and isolating rare events exhibiting non-linear correlations on the fly during experimental measurements, which can help steer exploration and improve the efficiency of experiments.
\end{abstract}

\maketitle
\section{Introduction}

Cold Target Recoil Ion Momentum Spectroscopy (COLTRIMS), often referred to as ``reaction microscopy'', has revolutionized our ability to study atomic and molecular dynamics by enabling kinematically complete measurements of multi-particle fragmentation processes.\cite{dorner_cold_2000,ullrich_recoil-ion_2003,jahnke_photoelectron_2007} COLTRIMS allows the measurement of the three-dimensional momentum vectors of all ions and electrons produced in fundamental ionizing reactions in gaseous atoms, molecules, and complexes, providing unprecedented insight into quantum mechanical processes, correlations, and coherences.\cite{schoffler_ultrafast_2008,weber_complete_2004}

The comprehensive nature of COLTRIMS measurements, while powerful, presents significant data analysis challenges. For instance, a typical experiment studying the photo double ionization of water resulting in three-body dissociation (H$^+$ + O$^+$ + H + 2e$^-$ or H$^+$ + H$^+$ + O + 2e$^-$), in which the 3D momentum of the neutral fragment was determined via momentum conservation, produces events characterized by 15 momentum components ($p_x$, $p_y$, $p_z$ for each of the five particles in each reaction), and the mass-to-charge ratio of each ion. When combined with derived physical quantities such as kinetic energy release (KER), relative angles between particles, electron energy sharing, and others, the effective dimensionality can easily exceed 50 features per event. Modern COLTRIMS experiments routinely generate datasets containing tens of millions of such high-dimensional events. These events are represented as rows in the list-mode file format, which records the events on a shot-by-shot basis. The features (observables and derived quantities) are stored in the columns of this large table.\cite{fehre_absolute_2018}

The current well-established analysis approaches for COLTRIMS data have significant limitations:

\begin{enumerate}
\item \textbf{Projection-based methods}: Researchers typically create one- or two-dimensional histograms of selected variables, potentially missing important multi-dimensional correlations that reveal reaction mechanisms.\cite{demekhin_exploring_2011}

\item \textbf{Sequential filtering}: Applying cuts on selected parameters can introduce bias and obscure unexpected patterns in the data.\cite{czasch_partial_2005}

\item \textbf{Manual feature engineering}: The analysis relies heavily on domain expertise to construct relevant observables, potentially missing non-intuitive relationships between observables or parameters.\cite{schmidt_spatial_2012}
\end{enumerate}

Recent advances in machine learning,\cite{lecun_deep_2015} particularly in dimensionality reduction and unsupervised clustering, offer promising solutions to these challenges.\cite{wang_scientific_2023} The application of machine learning to high-dimensional experimental data analysis has seen remarkable growth across multiple domains of experimental physics, chemistry, biology and engineering.\cite{chen_machine_2021,hu_deep_2025} In X-ray scattering and spectroscopy, variational autoencoders and other dimensionality reduction techniques have been successfully applied to learn latent representations of complex scattering patterns,\cite{chen_machine_2021} enabling researchers to identify similar structures through latent space clustering and to generate new structures through sampling and exploration of the learned manifold. Generative adversarial networks have been successfully applied to analyze ultrafast electron diffraction images, demonstrating the utility of deep learning approaches for extracting structural dynamics from high-dimensional experimental data where traditional analysis methods face ambiguity.\cite{daoud_novel_2023} For small-angle X-ray scattering (SAXS) data, variational autoencoders have been employed to visualize large datasets in low-dimensional latent spaces, enabling rapid capture of key features such as similarity among scattering patterns and structural evolution trends.\cite{zhao_visualization_2023} In protein structure analysis, autoencoders trained to compress 3D shape information into 200-dimensional latent spaces have been combined with genetic algorithms to build 3D models consistent with scattering data,\cite{he_model_2020} demonstrating the power of latent space optimization for inverse problems.

Recent efforts have focused on creating integrated web-based platforms for ML-driven analysis of experimental data at scientific user facilities. The MLExchange platform introduced a novel labeling pipeline that accelerates annotation of large scientific datasets using AI-guided tagging techniques, with interconnected graphical user interfaces for data reduction, classification, latent space exploration, and label assignment.\cite{chavez_machine-learning-driven_2025} This approach has proven instrumental for pattern recognition in X-ray scattering data, enabling scientists to label large datasets efficiently while maintaining the ability to train and fine-tune customizable ML models within the same pipeline. Mass spectrometry imaging, which shares with COLTRIMS the challenge of high-dimensional, multi-parameter datasets, has benefited from variational autoencoder approaches that learn nonlinear spectral manifolds to reveal biologically relevant clusters.\cite{abdelmoula_peak_2021} Convolutional autoencoders have proven particularly effective at aggregating features and preserving low-abundant signals in their latent space representations,\cite{bitto_enhancing_2024} addressing challenges similar to those faced in identifying rare fragmentation channels in COLTRIMS data.

The Uniform Manifold Approximation and Projection (UMAP) technique\cite{mcinnes_umap_2020} has emerged as a particularly powerful tool for dimensionality reduction in biological and physical sciences.\cite{becht_dimensionality_2019,szubert_structure-preserving_2019} Compared to other dimensionality reduction methods, UMAP provides high reproducibility, and meaningful organization of clusters,\cite{becht_dimensionality_2019} while preserving both local neighbor relations and aspects of global structure. Applications in molecular dynamics simulations of biomacromolecules have demonstrated UMAP's superior performance when compared with linear reduction methods and competitive performance with other nonlinear techniques.\cite{trozzi_umap_2021} In particle physics, machine learning has transformed data analysis and simulation, with applications ranging from boosted decision trees for classification to various types of neural networks for pattern recognition and event reconstruction.\cite{bourilkov_machine_2019,guest_deep_2018} Reinforcement learning approaches have been applied to hierarchical clustering problems in particle physics, demonstrating that such tasks can be phrased as Markov Decision Processes.\cite{brehmer_hierarchical_2020}

However, existing machine learning tools are typically not tailored to the specific requirements of tabulated momentum spectroscopy data, lacking physics-informed constraints, configurable molecular systems for different experimental targets, adaptive quality assessment for clustering reliability, and interactive exploration tools for multi-dimensional correlations. In parallel with the present work, Venkatachalam et al.~\cite{venkatachalam_exploiting_2025} recently reported on ML-assisted Coulomb explosion imaging (CEI) guided by classical simulations to extract molecular geometries and distinguish between isomers of mid-sized molecules from list-mode data. In another concurrent study, Li et al.~\cite{li_generative_2026} employed a deep generative neural network to infer molecular structures from ion momentum distributions generated during the rapid Coulomb explosion of molecules with up to nine atoms. In both approaches, the complete CEI, where all atomic ionic fragments of the molecule are detected, is currently a prerequisite, and based on this, the authors demonstrate its potential to investigate molecular transformation in real-time with unprecedented detail. Beyond complete CEI, which is well suited for strong-field laser and XFEL CEI experiments, the momentum-spectroscopy community will benefit from an integrated platform that unifies modern dimensionality-reduction methods with domain-specific analysis tools in an accessible, web-based environment capable of correlating electrons, ions, and at least one neutral fragment in a general and flexible way.

We present SCULPT, an integrated software platform that bridges this gap by combining state-of-the-art machine learning techniques with physics-informed analysis tools specifically designed for multi-particle coincidence list-mode data. The key innovations of SCULPT include:

\begin{enumerate}
\item \textbf{Physics-based dimensionality reduction}: The integration of UMAP\cite{mcinnes_umap_2020} with automated physics feature calculation ensures that dimensionality reduction preserves physically meaningful structures.

\item \textbf{Adaptive and quantitative reliability assessment}: A confidence scoring system that evaluates user-selected clustering quality metrics through weighted combination provides interpretable reliability assessments of UMAP results.

\item \textbf{Interactive exploration of multidimensional data by arbitrary dimensionality reduction}: Dynamic visualization with selection and filtering tools enables a hypothesis-driven analysis unrestricted by conventional 1-D or few-dimensional projections.

\item \textbf{Flexible configurable molecular systems}: A profile-based system for defining arbitrary molecular configurations calculates appropriate mass- and charge-dependent features.

\item \textbf{Automated feature discovery}: Genetic programming algorithms\cite{fortin_deap_2012} that discover mathematical combinations of input features aid in discovering relevant clusters and correlations.
\end{enumerate}

\section{Software Architecture and Implementation}
SCULPT is implemented as a web-based application using Python and Plotly Dash framework,\cite{plotly_technologies_inc_collaborative_2015} chosen for its ability to create reactive, interactive visualizations without requiring client-side programming. The platform follows a modular design that integrates data processing, machine learning analysis, quality assessment, and interactive visualization capabilities into a cohesive workflow for COLTRIMS data exploration.

\subsection{Data Processing and Physics Feature Calculation}

\begin{figure*}[htbp]
\centering
\includegraphics[width=2\columnwidth]{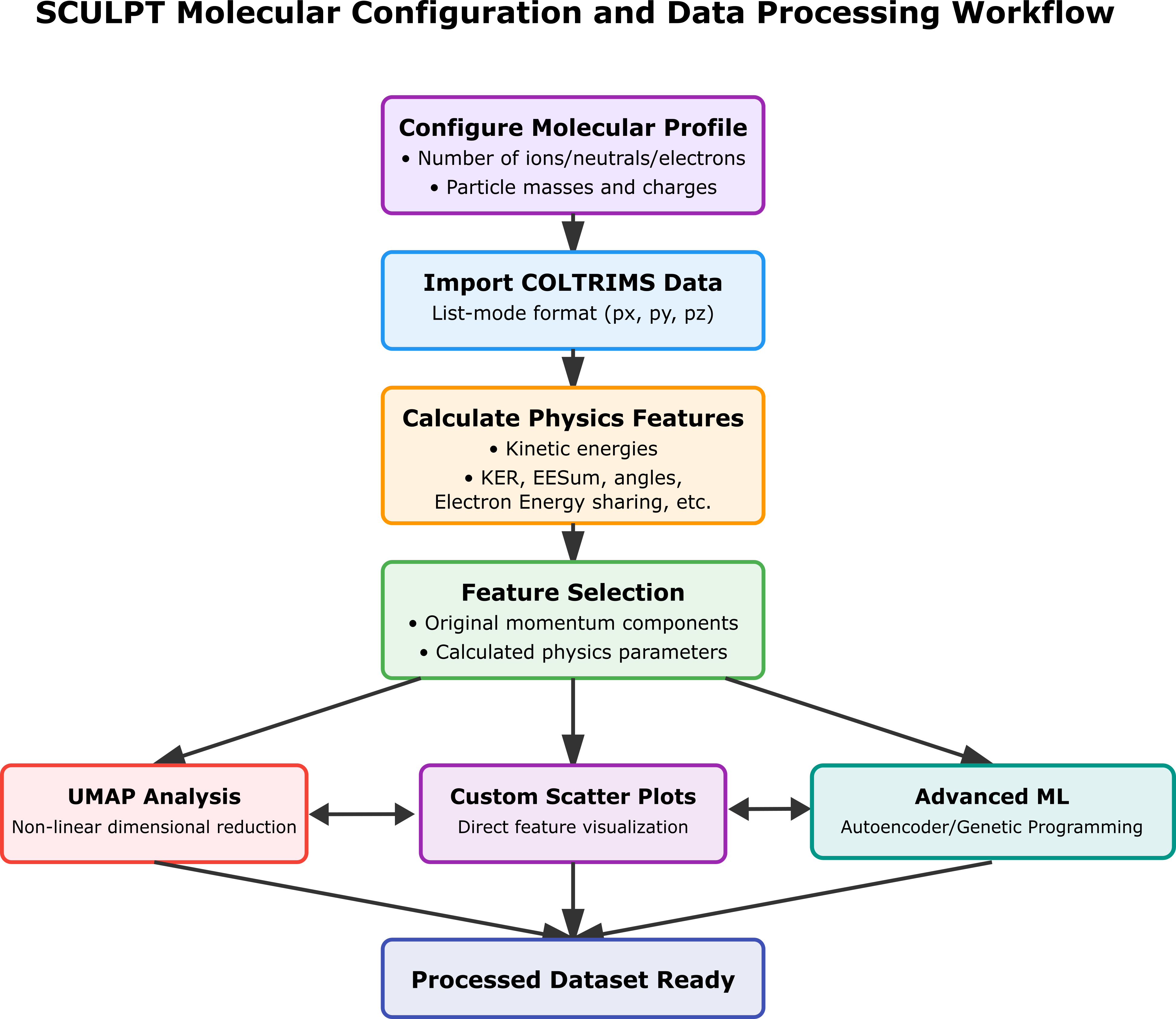}
\caption{Flowchart showing the molecular configuration and data processing workflow.}
\label{fig:molecular_profiles}
\end{figure*}

The data processing layer handles importing COLTRIMS data files and implements a novel molecular configuration system. Users can define molecular profiles by specifying particle types, masses, and charges (see Fig.~\ref{fig:molecular_profiles}). 

Moreover, it enables accurate physics calculations for various molecular photoionization experiments, with targets such as different isotopologues (e.g., H$_2$O, D$_2$O, HDO), as well as other molecular systems, without source code modification.

The physics feature calculator automatically computes:
\begin{itemize}
\item Individual particle kinetic energies: $E_i = p_i^2/(2m_i)$
\item Kinetic Energy Release of dissociation processes: $\text{KER} = \sum E_{\text{ions}}$
\item Sum of electron energies: $\text{EESum} = \sum E_{\text{electrons}}$
\item Relative angles between all particle pairs
\item Energy sharing parameters such as electron energy sharing: $\text{EESharing} = E_{\text{electron}}/\sum E_{\text{electrons}}$ 
\end{itemize}

\subsection{Machine Learning Components}

\subsubsection{UMAP Implementation}

SCULPT employs UMAP\cite{mcinnes_umap_2020} for non-linear dimensionality reduction of high-dimensional COLTRIMS data. UMAP constructs a topological representation of the high-dimensional data by modeling each data point's relationship to its nearest min, then optimizes a low-dimensional embedding that preserves these local neighborhood structures while maintaining global data topology. This approach is particularly effective for COLTRIMS data as it reveals non-linear correlations between momentum components and derived physics quantities that are not apparent in traditional linear projection methods. The user selects a set of calculated physics features (kinetic energies, angular distributions, correlation parameters) or original momentum components as the inputs to be embedded into the UMAP algorithm. The implementation preserves file labels throughout the dimensionality reduction process, enabling direct comparison of multiple datasets or experimental conditions within the same 2D visualization space. Key UMAP parameters including the number of neighbors ($n_{\text{neighbors}}$, controlling local versus global structure preservation) and minimum distance ($\text{min\_dist}$, controlling cluster compactness) are user-configurable, with typical values of $n_{\text{neighbors}}=15$ and $\text{min\_dist}=0.1$ providing effective visualizations for COLTRIMS datasets.
\subsubsection{Deep Autoencoder Architecture}

The autoencoder module implements a symmetric neural network architecture using PyTorch for feature learning from high-dimensional COLTRIMS data. The network consists of an encoder that progressively compresses the input feature space through multiple hidden layers to a low-dimensional latent representation, and a symmetric decoder that reconstructs the original features from this compressed representation. The latent space dimensionality (typically 2-10 dimensions) is user-configurable, with the compression forcing the network to learn the most salient features that capture the essential variance in the fragmentation dynamics. Training employs mean squared error reconstruction loss with the Adam optimizer, and the resulting latent features can be visualized directly or used as input to subsequent UMAP projection and clustering analysis. This approach is particularly effective for discovering non-linear correlations between momentum components and derived physics quantities that may not be apparent through traditional feature selection methods.

\subsubsection{Genetic Programming for Feature Discovery}

SCULPT implements genetic programming using symbolic regression to discover interpretable feature combinations.\cite{fortin_deap_2012,cherrier_consistent_2019} Genetic programming has proven effective for automated feature construction in scientific data analysis, particularly when the goal is to discover non-trivial mathematical relationships between input variables that may not be apparent through traditional feature engineering.\cite{lacava_learning_2020,makke_interpretable_2024} This approach is explicitly chosen to discover non-trivial features that might provide superior cluster separation compared to standard physics parameters (like energy and momentum conservation). The method has been successfully applied to experimental physics data, where dimensional consistency and physical interpretability of constructed features are critical requirements.\cite{cherrier_consistent_2019} By evolving populations of mathematical expressions through selection, crossover, and mutation operations, the genetic programming module can identify complex combinations of momentum components, energies, and angular features that capture underlying correlations in fragmentation patterns, thereby enhancing the ability to distinguish between different quantum states or reaction pathways in COLTRIMS data.

\subsection{Adaptive Quality Assessment}
A key innovation in SCULPT is the adaptive confidence scoring system that provides quantitative reliability assessments for clustering results. The system evaluates user-selected clustering quality metrics, each automatically assigned predefined weights and reliability scores based on their robustness and general applicability. Users select which metrics to calculate through the interface, allowing flexibility and balance between computational cost and analysis depth. The confidence score is then computed as a weighted combination of only the selected metrics, with weights normalized to account for that subset of metrics. Metrics are organized into three tiers based on their general reliability: Tier 1 metrics (silhouette score, Hopkins statistic) are highly reliable and recommended for all analyses; Tier 2 metrics (cluster stability, physics consistency, Calinski-Harabasz index) provide additional validation with moderate reliability; and Tier 3 metrics (Davies-Bouldin index) are included with heavily reduced weight due to known limitations. The only automatic exclusion occurs for the physics consistency metric, which is omitted when physics features are unavailable in the dataset. These reliability tiers and associated weights were empirically determined through validation on the D$_{2}$O double ionization dataset with known ground truth quantum state assignments.\cite{reedy_dissociation_2018} The tier assignments reflect performance patterns observed when separating the eight distinct dication states in this well-characterized system, where the metrics' ability to correctly identify physically meaningful clusters could be verified against established quantum state separations. 
\textcolor{black}{The adaptive confidence score is intended as a heuristic guide for exploratory analysis rather than a rigorous statistical measure of physical correctness. In practice, scores are categorized as follows: $\ge$ 0.65 (High) indicates reliable cluster separation suitable for drawing preliminary conclusions; 0.50–0.65 (Moderate) suggests reasonably reliable results that warrant further validation; 0.35–0.50 (Low) indicates results should be used with caution and may benefit from alternative feature selection or UMAP parameters; and $\le$ 0.35 (Very Low) suggests the clustering may not capture meaningful structure.}

\textbf{Tier 1 metrics} (highest reliability, recommended for all analyses):
\begin{enumerate}
\item \textbf{Silhouette Score} $S \in [-1, 1]$: This metric measures both cluster cohesion (how similar a point is to others in its own cluster) and separation (how dissimilar it is to points in neighboring clusters).\cite{rousseeuw_silhouettes_1987} For each data point $i$, the silhouette coefficient is calculated as $s(i) = \frac{b(i) - a(i)}{\max\{a(i), b(i)\}}$, where $a(i)$ is the mean distance to other points in the same cluster and $b(i)$ is the mean distance to points in the nearest neighboring cluster. The overall silhouette score is the mean across all points. Values near +1 indicate well-separated clusters, values near 0 suggest overlapping clusters, and negative values indicate potential misclassification. This metric receives a weight of $w=0.35$ (contributing 35\% to the final confidence score) and a reliability score of $r=0.9$ (indicating 90\% confidence in its assessment validity). The weight $w$ is the contribution to the overall confidence score, and $r$ represents the reliability or confidence in the validity of the metric. The weights sum to unity across all active metrics, and reliability scores modulate the influence of each metric based on empirical performance across diverse clustering scenarios.

\item \textbf{Hopkins Statistic} $H \in [0, 1]$: This metric assesses the clustering tendency of the dataset by measuring the probability that the data is generated from a uniform distribution versus containing meaningful clusters.\cite{hopkins_new_1954,lawson_new_1990} The statistic compares the nearest-neighbor distances for a sample of real data points against those for randomly generated points within the same space. Values of $H > 0.75$ indicate strong clusterability (the data significantly deviates from uniform randomness), while $H < 0.5$ suggests the data may be uniformly distributed with no inherent cluster structure. This metric receives $w=0.25$ and $r=0.85$, reflecting its importance in determining whether clustering is appropriate for the dataset. Here, $H$ refers specifically to the Hopkins statistic value. 
\end{enumerate}

\textbf{Tier 2 metrics} (moderate reliability, useful for comprehensive assessment):
\begin{enumerate}
\item \textbf{Cluster Stability} $\in [0, 1]$: Measures the reproducibility of cluster assignments under small perturbations by injecting Gaussian noise (5\% of feature standard deviation) into the data and comparing the resulting cluster assignments with the original using the Adjusted Rand Index.\cite{hubert_comparing_1985} High stability values ($>0.8$) indicate robust clusters that are insensitive to minor variations in the data, while low values suggest the clustering may be capturing noise rather than genuine structure. This metric receives $w=0.15$ and $r=0.7$.

\item \textbf{Physics Consistency} $\in [0, 1]$: This metric is automatically 
excluded when physics features are unavailable in the dataset (e.g., when 
analyzing autoencoder latent representations). It is a domain-specific validation metric that measures the ratio of within-cluster variance to between-cluster variance for key physics parameters. The metric evaluates a predefined set of physics quantities: KER, EESum, EESharing, individual particle energies (ions and electrons), and total energy. For each available physics parameter, the metric calculates $PC_i = \frac{\text{Var}_{\text{between}}}{\text{Var}_{\text{between}} + \text{Var}_{\text{within}}}$, where $\text{Var}_{\text{between}}$ is the variance of cluster means weighted by cluster size, and $\text{Var}_{\text{within}}$ is the weighted average of within-cluster variances. The overall physics consistency is the mean across all evaluated parameters. Values near 1 indicate that clusters correspond to physically distinct states with homogeneous physics parameter distributions within each cluster, even if these physics parameters were not explicitly used as clustering features. This provides an independent validation that the clustering captures meaningful physical differences rather than arbitrary data structure. This metric receives $w=0.2$ and $r=0.8$.

\item \textbf{Calinski-Harabasz Index} $\in [0, \infty)$: Measures cluster separation by computing the ratio of between-cluster dispersion to within-cluster dispersion in the UMAP latent space.\cite{calinski_dendrite_1974} The index is defined as $CH = \frac{\text{tr}(B_k)}{\text{tr}(W_k)} \times \frac{N - k}{k - 1}$, where $B_k$ is the between-cluster dispersion matrix, $W_k$ is the within-cluster dispersion matrix, $N$ is the number of data points, and $k$ is the number of clusters. Higher values indicate better-defined clusters with greater separation in the reduced-dimensional space. This metric receives $w=0.1$ and $r=0.6$.
\end{enumerate}

\textbf{Tier 3 metrics} (lower reliability, heavily downweighted):
\begin{enumerate}
\item \textbf{Davies-Bouldin Index} $\in [0, \infty)$: Measures the average similarity ratio between each cluster and its most similar neighbor, where similarity is defined as the ratio of within-cluster scatter to between-cluster separation.\cite{davies_cluster_1979} Lower values indicate better clustering. However, this metric is known to be sensitive to cluster shape assumptions (favoring spherical clusters) and can produce misleading results for elongated or irregular cluster geometries common in UMAP embeddings.\cite{arbelaitz_extensive_2013} Due to these limitations, it receives minimal weight ($w=0.005$) and low reliability ($r=0.4$).
\end{enumerate}

Additionally, the system tracks the \textbf{Noise Ratio} $\in [0, 1]$, representing the fraction of points classified as noise by the clustering algorithm. SCULPT implements DBSCAN (Density-Based Spatial Clustering of Applications with Noise)\cite{ester_density-based_1996} as the primary clustering algorithm, which automatically identifies noise points as those not belonging to any dense region. For the main UMAP analysis and custom scatter plots, DBSCAN uses automatic parameter optimization where the algorithm searches for the optimal epsilon value (distance threshold) while maintaining a fixed minimum sample requirement of 5 neighbors. The optimization procedure tests epsilon values in the range $[0.1, 1.0]$ and selects the configuration that maximizes the number of clusters while keeping the noise ratio below 50\%. In the advanced genetic programming module, users have full control over both the epsilon and minimum samples parameters, allowing for fine-tuning based on specific dataset characteristics. Points are labeled as noise if they have fewer than the minimum number of neighbors within the specified radius. The noise ratio influences both reliability assessment and performance bonuses, with very low noise ratios ($<0.05$) indicating clean cluster separation.

The adaptive confidence score $C$ serves as a summary metric that combines all individual quality assessments into a single interpretable value. It employs a reliability-weighted calculation:

\begin{equation}
C = \frac{\sum_{i} w_i \cdot r_i \cdot n_i}{\sum_{i} w_i \cdot r_i},
\end{equation}

where $w_i$ is the weight for metric $i$, $r_i$ is its reliability score, and $n_i$ is the normalized value of metric $i$. The sum ranges only 
over the metrics selected by the user for calculation. The normalization function $n_i$ transforms each metric to a $[0,1]$ scale using metric-specific scaling functions designed to provide realistic assessments across the full range of possible values.

The system incorporates several advanced features to ensure robust assessment:

\textbf{Adaptive normalization} employs metric-specific scaling functions that account for the different value ranges and distributions of each metric. For the silhouette score, the transformation is $n_S = 0.5(S + 1)$, mapping $[-1, 1]$ to $[0, 1]$. For the Hopkins statistic, $n_H = H$ (already in $[0,1]$). For the Calinski-Harabasz index, a logarithmic transformation $n_{CH} = \min(1, \log(1 + CH)/\log(1000))$ prevents extremely large values from dominating. For the Davies-Bouldin index, $n_{DB} = \max(0, 1 - DB/3)$ transforms lower-is-better to higher-is-better with appropriate scaling. These functions avoid overly harsh penalties for moderate clustering while preserving sensitivity to exceptional results, with normalized values capped at 0.98 to reserve perfect scores for theoretical limits.

\textbf{Critical threshold management} applies contextual penalties to prevent overconfident assessments of poor clusterings. Severe failures (silhouette score $S < -0.1$ or Hopkins statistic $H < 0.3$) cap the confidence at 0.4, indicating fundamental clustering problems. Borderline cases are identified as those with $0.2 < S < 0.4$ or $0.5 < H < 0.7$, which receive moderate penalties (confidence capped at $\leq 0.7$) to reflect ambiguous cluster quality.

\textbf{Performance bonuses} are applied using asymptotic scaling to reward exceptional clustering without inflating confidence unrealistically. Four bonus conditions are evaluated: exceptional silhouette scores ($S > 0.6$, $+0.1$ bonus), very low noise ratios (noise $< 0.05$, $+0.05$ bonus), high stability (stability $> 0.8$, $+0.05$ bonus), and strong clustering tendency ($H > 0.8$, $+0.05$ bonus). The bonus system uses asymptotic scaling: $C_{\text{new}} = C + (0.95 - C) \cdot \frac{\text{bonus}}{0.95}$ to maintain realistic confidence bounds and prevent scores from approaching 1.0.

\textbf{Uncertainty quantification} provides confidence intervals around the final score to communicate the reliability of the assessment. The base uncertainty is calculated as $\sigma_{\text{base}} = 0.1 + \max(0, (3 - n_{\text{metrics}})) \times 0.05$, where $n_{\text{metrics}}$ is the number of active metrics used in the calculation. This function was empirically derived through iterative testing on the eight known dication states in the D$_{2}$O double ionization dataset,\cite{reedy_dissociation_2018} where the uncertainty model parameters (0.1 baseline and 0.05 penalty per missing metric) were tuned to reflect observed variations in metric reliability. The 0.1 baseline represents the estimated inherent measurement uncertainty, while the penalty term increases uncertainty when fewer metrics are available (reflecting reduced information). The base uncertainty is further adjusted for the average reliability of active metrics: $\sigma_{\text{total}} = \sigma_{\text{base}} + \max(0, (0.8 - r_{\text{avg}})) \times 0.1$, where $r_{\text{avg}} = \frac{\sum_i r_i}{n_{\text{metrics}}}$. This adjustment increases uncertainty when relying on less reliable metrics.

\subsection{Advanced Data Filtering and Selection}

SCULPT incorporates sophisticated data filtering capabilities that enable users to refine datasets at multiple stages of the analysis pipeline, enhancing both computational efficiency and analytical focus. The system provides three primary filtering mechanisms: (1) feature-based selection, (2) density-based filtering, and (3) physics parameter filtering.

\subsubsection{Feature Selection Framework}

The feature selection system organizes available variables into hierarchical categories based on their physical significance and computational origin. Features are automatically categorized as follows:

\textbf{Original Momentum Components:} Raw momentum measurements ($p_x$, $p_y$, $p_z$) for each detected particle, providing the fundamental kinematic information from experimental data.

\textbf{Momentum Magnitudes:} Derived scalar quantities $|\vec{p}_i| = \sqrt{p_{x,i}^2 + p_{y,i}^2 + p_{z,i}^2}$ for each particle, offering magnitude-based clustering without directional bias.

\textbf{Energy Variables:} Comprehensive energy features including individual particle kinetic energies, Kinetic Energy Release representing the sum of ion energies, electron energy sum, electron energy sharing ratios, and total system energy.

\textbf{Angular Features:} Laboratory-frame polar angles $\theta_i = \arccos(p_{z,i}/|\vec{p}_i|)$ and azimuthal angles $\phi_i = \arctan2(p_{y,i}, p_{x,i})$ for each particle, enabling analysis of angular distributions and correlations.

\textbf{Correlation Features:} Inter-particle relationships including dot products $\vec{p}_i \cdot \vec{p}_j$, momentum differences, and angles between particle pairs $\cos(\alpha_{ij}) = \frac{\vec{p}_i \cdot \vec{p}_j}{|\vec{p}_i||\vec{p}_j|}$.

The system provides hierarchical selection controls with category-level selection and individual feature toggling, which enables efficient exploration of the high-dimensional feature spaces, while maintaining the interpretability of the selected subset.

\subsubsection{Density-Based Filtering}

Density-based filtering addresses the challenge of computational scalability and noise reduction by selectively retaining high-density regions in both UMAP projections and direct feature scatter plots. The implementation uses efficient grid-based density estimation that operates on any two-dimensional data representation:

\begin{equation}
\rho(x_i, y_i) = \frac{1}{N} \sum_{j=1}^{N} G_{\sigma}(x_i - x_j, y_i - y_j),
\end{equation}

where $G_{\sigma}$ is a Gaussian kernel with bandwidth parameter $\sigma$ (user-adjustable from 0.01 to 1.0), and $(x_i, y_i)$ represents coordinates in either the 2D UMAP embedding space or direct feature space (e.g., energy vs. angle scatter plots). The system creates a 100×100 grid histogram of the selected coordinates, applies Gaussian smoothing with $\sigma_{smooth} = 10\sigma$, and assigns density values to each point based on grid interpolation.

Points are filtered using a percentile-based threshold $\tau_p$, where users specify the percentile (0-100\%) of density values to retain. Points with density $\rho \geq \tau_p$ are preserved, effectively removing sparse outliers while maintaining the core cluster structure. This approach significantly reduces computational overhead for subsequent analyses, while preserving essential clustering patterns.

For scatter plot applications, density filtering enables users to focus on statistically significant regions within the original feature space by removing experimental noise and low-statistics outliers that might otherwise obscure underlying physical correlations. The same mathematical framework applies whether filtering UMAP embeddings or direct two-dimensional projections of calculated physics parameters.

\subsubsection{Physics Parameter Filtering}

Physics parameter filtering enables domain-specific data selection based on physically meaningful quantities. The system supports filtering on commonly used physics parameters including kinetic energy release (KER), individual particle kinetic energies for ions and electrons, electron energy sums and energy sharing ratios, inter-particle angular correlations, and momentum differences between particles. This filtering capability enables selection of specific energy regimes, geometric configurations, or correlation patterns corresponding to different physical processes.

For each selected parameter, the system automatically determines the data range and provides an interactive range slider with intelligent binning. The parameter range is dynamically calculated as $[\lfloor \min(P) \rfloor, \lceil \max(P) \rceil]$ where $P$ represents the selected parameter values across all loaded datasets. The filtering condition retains data points where $P_{\min} \leq P_i \leq P_{\max}$, based on user-specified bounds.

\subsubsection{Integrated Filtering Pipeline}

The filtering system operates through an integrated pipeline where filters can be applied sequentially:

1. \textbf{Feature Selection:} Users first select relevant features for dimensional reduction, reducing computational complexity and focusing on scientifically relevant dimensions.

2. \textbf{UMAP Projection:} The selected features undergo UMAP embedding to create the 2D visualization space.

3. \textbf{Density Filtering:} Optional density-based filtering removes low-density outliers from the UMAP space, improving signal-to-noise ratio.

4. \textbf{Physics Filtering:} Optional parameter-based filtering selects physically meaningful subsets for targeted analysis.

Each filtering step preserves data provenance, maintaining links between filtered subsets and original experimental data. The system provides detailed filtering statistics, which includes retention percentages and per-file event counts, enabling users to assess the impact of each filtering operation.

This multi-stage filtering approach enables efficient exploration of large experimental datasets while maintaining physical interpretability and statistical rigor, supporting both exploratory data analysis and hypothesis-driven investigation of specific reaction mechanisms.

\subsection{Interactive Visualization and Selection Tools}

SCULPT provides multiple visualization modes with consistent interaction patterns:

\begin{enumerate}
\item \textbf{Scatter plots}: Interactive 2D projections with lasso and box selection tools
\item \textbf{Density heatmaps}: Kernel density estimation for identifying high-density regions
\end{enumerate}

Selection tools are implemented using Plotly's built-in selection capabilities. These tools empower the user to identify and isolate groups of events and save them in separate files for further processing. To further understand the physics that correlates events in any one cluster, SCULPT enables the user to visualize events in scatter plots or heat maps according to any of the original features such as total energy, kinetic energy release (KER), etc. This allows for vetting the data, visualizing the clusters identified by SCULPT on lower-dimensional phase spaces and understanding the physics behind the clustering.

\textcolor{black}{
\subsection{Design Philosophy and Default Parameters}
In this section we shed light on the default parameter choices implemented in SCULPT.\newline
\textbf{Confidence score weighting:}
The tiered metric system emphasizes local density structure (silhouette score, weight 0.35) and intrinsic clusterability (Hopkins statistic, weight 0.25) because these properties are most relevant for identifying physically distinct fragmentation channels in momentum space. Metrics known to be sensitive to cluster shape assumptions, such as the Davies-Bouldin index, are heavily downweighted (weight 0.005) since UMAP embeddings frequently produce non-spherical cluster geometries. This weighting scheme may undervalue clustering quality in datasets where global structure is more important than local density, or where clusters have highly uniform, spherical shapes.\newline
\textbf{DBSCAN parameter selection:}
The automatic epsilon optimization maximizes the number of detected clusters subject to a 50\% noise threshold. This heuristic favors segmentation—revealing fine structure that might otherwise be merged—over conservative clustering that minimizes false positives. For applications requiring high-purity cluster assignments, users should consider manually adjusting epsilon toward larger values or using the advanced analysis module where full control over DBSCAN parameters is available.\newline
\textbf{UMAP parameters:}  
The default values ($n_{\text{neighbors}}=15$, $\text{min\_dist}=0.1$) balance local and global structure preservation for typical COLTRIMS dataset sizes (10$^{4}$–10$^{6}$ events). Smaller $n_{\text{neighbors}}$ values emphasize local structure and may reveal finer sub-clustering, while larger values preserve more global topology at the cost of local detail. These parameters should be adjusted based on dataset size and the scale of structure being investigated.\newline
The current defaults may perform poorly for datasets with highly imbalanced cluster sizes, where small but physically important clusters may be classified as noise; for data with continuous gradients rather than discrete clusters; or for very small datasets ($<$ 1000 events) where the Hopkins statistic and stability metrics become unreliable. In such cases, users should rely more heavily on visual inspection and physics-based validation rather than the confidence score alone. Additionally, the filtering features incorporated in SCULPT serve to aid users identify clusters that would otherwise be challenging to isolate.
}

\section{User Interface and Workflow}

\subsection{Interface Organization}

\begin{figure*}[htbp]
\centering
\includegraphics[width=2\columnwidth]{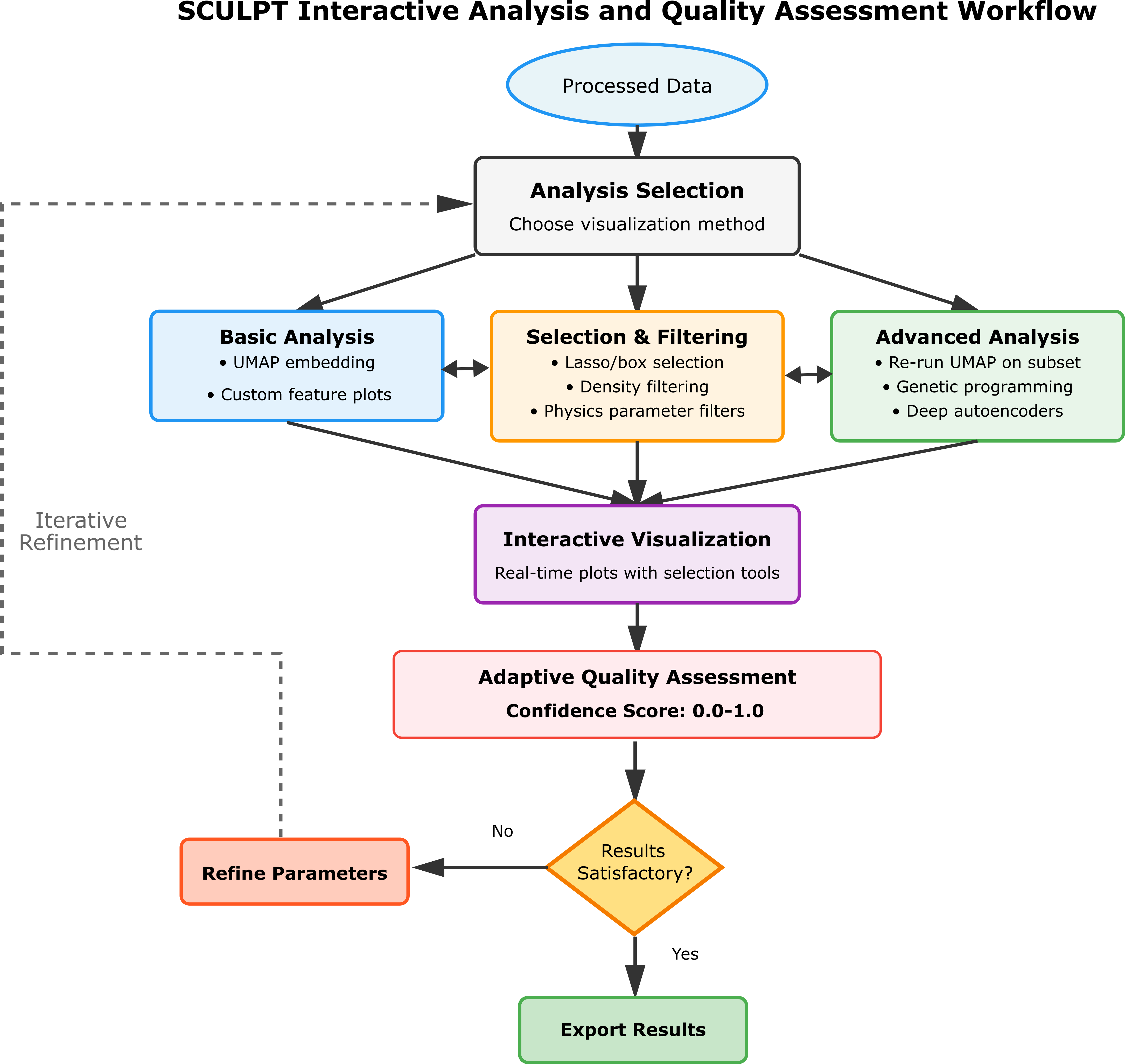}
\caption{Flowchart showing the interactive analysis and quality assessment workflow.}
\label{fig:interface}
\end{figure*}

SCULPT's interface is organized into five main sections (see Fig.~\ref{fig:interface}):

\begin{enumerate}
\item \textbf{Data \& Configuration}: File upload and molecular profile management
\item \textbf{Basic Analysis}: Probabilistic machine learning (UMAP embedding) and custom feature plots
\item \textbf{Selection \& Filtering}: Interactive data selection and filtering tools
\item \textbf{Advanced Analysis}: Re-analysis of selected subsets
\item \textbf{Machine Learning}: Autoencoder and genetic programming modules
\end{enumerate}

\subsection{Typical Analysis Workflow}

A typical analysis workflow proceeds as follows:

\begin{enumerate}
\item \textbf{Data Import}: Upload COLTRIMS data files and select appropriate molecular configuration
\item \textbf{Initial Exploration}: Run UMAP to visualize data structure.
\item \textbf{Cluster Identification}: Use DBSCAN with automatic parameter optimization or just rely on manual data separation by identifying clusters visually.
\item \textbf{Quality Assessment}: Review confidence scores and individual metrics.
\item \textbf{Refinement}: Select interesting regions for detailed analysis.
\item \textbf{Feature Discovery}: Optionally, apply genetic programming to attempt to discover new correlations.
\item \textbf{Validation}: Verify results by understanding the underlying physics, for example via plotting the identified clusters on physics-informed phase spaces.
\item \textbf{Export}: Save filtered datasets and discovered features as labeled list-mode data for further processing.
\end{enumerate}
All these steps are to be reiterated with the data in saved files for further processing.

\section{Case Study: D$_2$O Double Ionization}\label{section:casestudy}

To demonstrate SCULPT's capabilities with a well-conditioned example, we present a reanalysis of the D$_2$O double ionization following single photon absorption at 61 eV. Particularly, we considered the D$_2$O$^{2+}$ $\rightarrow$ D$^+$ + D$^+$ + O dissociation channel that was previously analyzed \textcolor{black}{for H$_2$O} by Reedy et. al.\cite{reedy_dissociation_2018} We created ground truth data based on the analysis in that work to isolate each of the quantum states of the water dication following dissociative photo- double ionization. This resulted in eight event list data files, containing the 3D momentum vector components of each particle for each event, for each quantum state: $^{3}A_{2}$ ,$^{3}B_{1}$, $^{3}B_{2}$ (producing oxygen O $^{3}$P),1$^{1}A_{1}$, 2$^{1}A_{1}$, $^{1}B_{1}$, $^{1}B_{2}$ (producing O $^{1}$D), 3$^{1}A_{1}$ (producing O $^1$S). 

\subsection{Experimental Data}

The dataset, consisting of the eight aforementioned files, contains around 1,900,000 coincidence events where four of the five particles were detected: D$^+$ + D$^+$ + O + e$^-$ + e$^-$. Each event is characterized by 15 momentum components (p$_{x}$, p$_{y}$, p$_{z}$ for each of the five particles). \textcolor{black}{For computational efficiency, all of the UMAP analyses presented here were conducted on a random 1\% sample of the full dataset. This sampling approach allows for rapid iterative exploration while preserving the essential clustering structure of the data. The sample size is chosen to provide reasonably low statistical uncertainties and sufficient density over the measured phase space to visualize clusters. After identifying clusters of interest on a sampled subset, we recommend users to re-run UMAP with different sampling fractions (ideally up to the full dataset) to verify that the cluster structure is preserved.}

\subsection{Analysis Results}
\subsubsection{UMAP Visualization}

\begin{figure*}[htbp]
\centering
\includegraphics[width=0.9\textwidth]{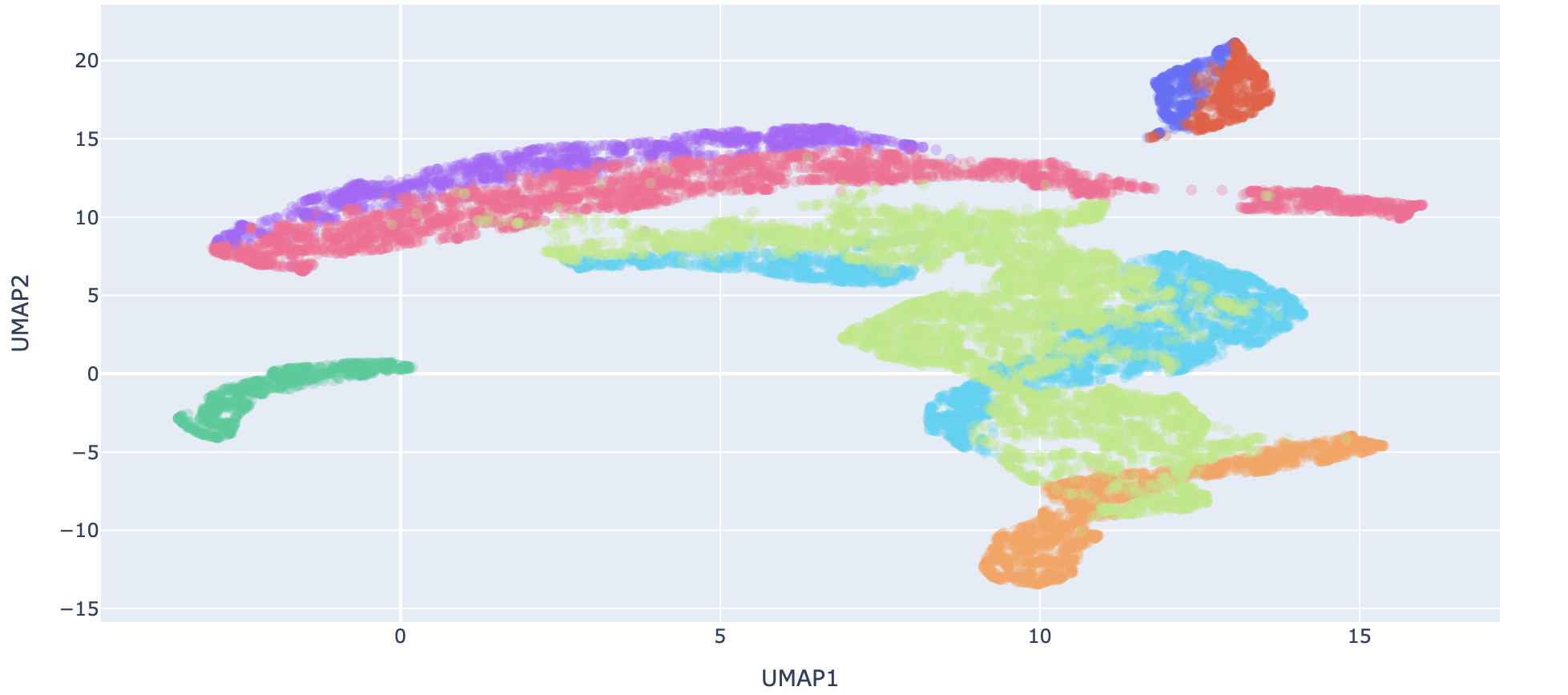}
\caption{UMAP projection of D$_2$O double ionization data showing five identified clusters. The calculated confidence score is 0.71. Colors represent different dication states and are for visualization purposes only.}
\label{fig:umap_d2o_A}
\end{figure*}

\begin{figure}[htbp]
\centering
\includegraphics[width=\columnwidth]{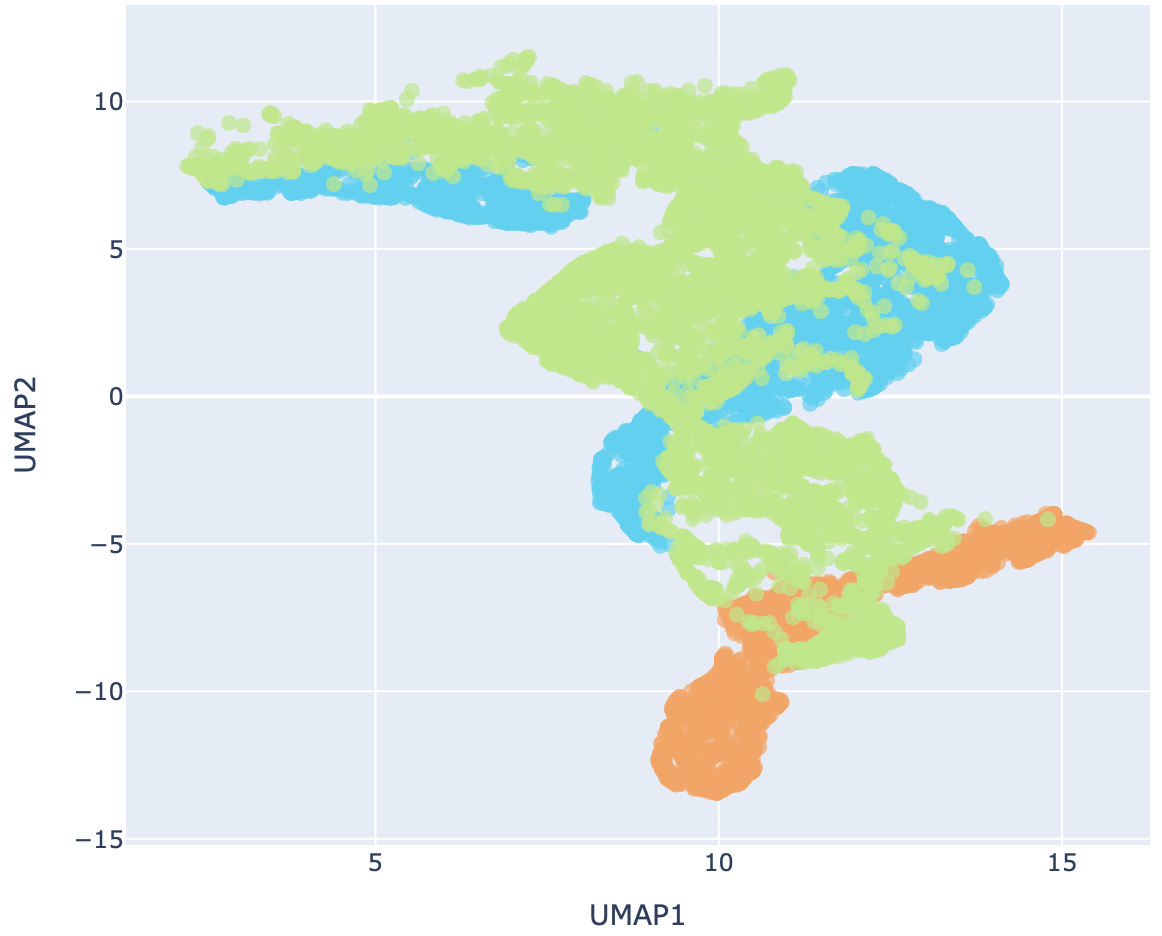}
\caption{Selected cluster containing three dication states. Colors represent different dication states and are for visualization purposes only.}
\label{fig:umap_d2o_B}
\end{figure}

The initial UMAP analysis revealed five distinct clusters, separated by white space in the 2D projection, as shown in Fig.~\ref{fig:umap_d2o_A}.
This UMAP analysis was run by selecting the features KER, EESum, Total Energy (KER + EESum), and the angle between $\alpha_{12}$ between the ion 1 and ion 2 momenta. Two of the clusters consist of one quantum state exclusively, while the other three clusters include two or more quantum states. These clusters need to be further analyzed to isolate each of the quantum states contained in them. Typically, the user would attempt to separate quantum states within large clusters while being guided by the confidence score and the visual check of achieving clear separations in UMAP plots. As an example, in Fig.~\ref{fig:umap_d2o_B}, we show the cluster containing the three states that was separated with the Lasso selection tool.

\begin{figure}[htbp]
\centering
\includegraphics[width=\columnwidth]{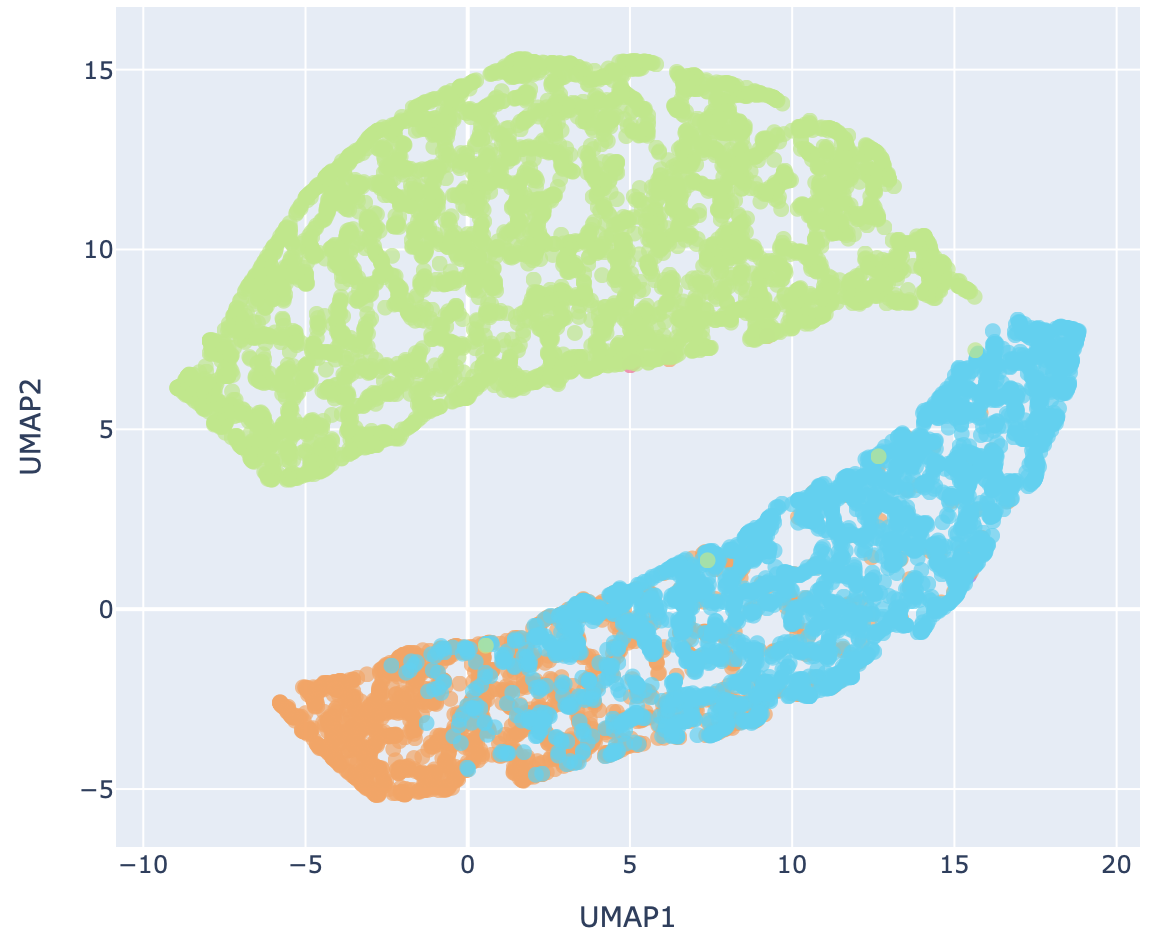}
\caption{UMAP projection of D$_2$O double ionization data of the selected cluster. Separation into two clusters is evident. The bottom cluster contains data from two dication states. The calculated confidence score is 0.70. Colors represent different dication states and are for visualization purposes only.}
\label{fig:umap_d2o_C}
\end{figure}

\begin{figure}[htbp]
\centering
\includegraphics[width=\columnwidth]{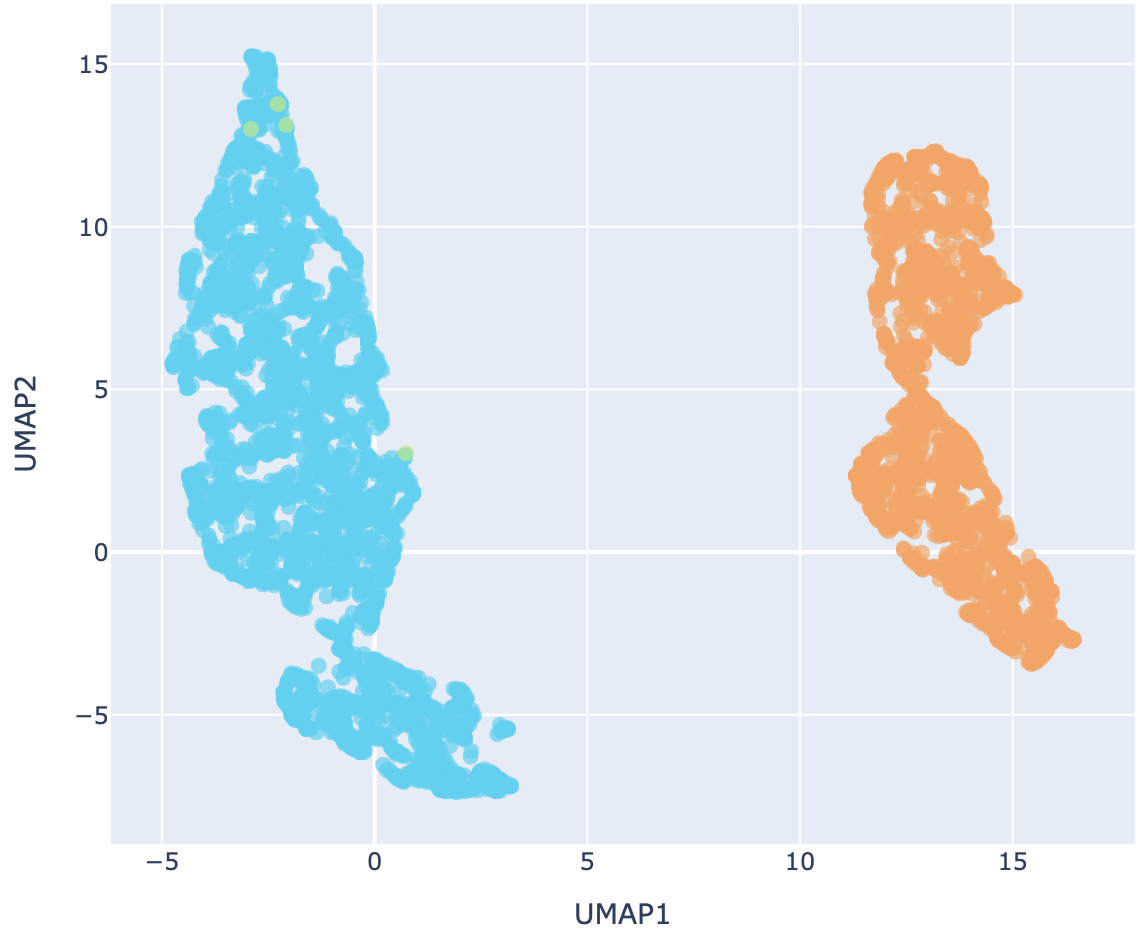}
\caption{UMAP projection of the lower cluster in Fig.~\ref{fig:umap_d2o_C}. Separation of the data into two dication states is achieved. The calculated confidence score is 0.79. Colors represent different dication states and are for visualization purposes only.}
\label{fig:umap_d2o_D}
\end{figure}

After isolating this cluster, we run UMAP on it again, this time while selecting the features KER, EESum, and Total Energy. The result, depicted in Fig.~\ref{fig:umap_d2o_C}, shows that the group is now separated into two distinct clusters. One of the clusters contains the two states producing the oxygen atom in the $^{1}$D state, specifically the water dication states $^{1}B_{2}$ (light-blue in Fig.~\ref{fig:umap_d2o_C}) and 2$^{1}A_{1}$ (orange). The calculated confidence score is 0.70 confirming a high confidence in separating the data according to the selected features.
We select the cluster containing the two states and run UMAP analysis on it, again using the features KER, EESum, Total Energy, and $\alpha_{12}$. The cluster is subsequently separated into two distinct clusters, each containing a single quantum state, as shown in Fig.~\ref{fig:umap_d2o_D}. A high confidence score of 0.79 is calculated.
Following the same general procedures described for the examples above, all eight quantum states in the present data are separable into individual clusters.

\subsubsection{Cluster Identification and Physics Interpretation}

A detailed analysis of the data enabled the isolation of the following clusters stemming from the D$_2$O$^{2+}$ $\rightarrow$ D$^+$ + D$^+$ + O fragmentation channel:

\textbf{Cluster 1} (dark-blue in Fig.~\ref{fig:umap_d2o_A}, 2.6\% of events): D$_2$O$^{2+}$ ($^{1}B_{1}$) dissociating to D$^{+}$ + D$^{+}$ + O($^{1}$D) 
\begin{itemize}
\item A singlet dication state that dissociates with a peak KER of $\sim$ 4.3 eV and a $\beta$ of $\sim$ 149$^{\circ}$. The kinematic signatures of this state overlap with those of the triplet dication states $^{3}B_{1}$ and 3$^{1}A_{1}$, making clean separation more challenging than some other examples presented here.
\end{itemize}

\textbf{Cluster 2} (red in Fig.~\ref{fig:umap_d2o_A}, 3.0\% of events): D$_2$O$^{2+}$ ($^{3}B_{1}$) dissociating to D$^{+}$ + D$^{+}$ + O($^{3}$P) 
\begin{itemize}
\item A triplet dication state with a peak KER of $\sim$ 4.3 eV and a $\beta$ of $\sim$ 146$^{\circ}$. It is characterized by a narrow $\beta$ distribution.
\end{itemize}

\textbf{Cluster 3} (dark-green in Fig.~\ref{fig:umap_d2o_A}, 4.6\% of events): D$_2$O$^{2+}$ (3$^{1}A_{1}$) dissociating to D$^{+}$ + D$^{+}$ + O($^{1}$S) 
\begin{itemize}
\item The only dication state leading to the O($^{1}$S) asymptote. It is a singlet state with a peak KER of $\sim$ 11.4 eV and a $\beta$ of $\sim$ 111$^{\circ}$.
\end{itemize}

\textbf{Cluster 4} (purple in Fig.~\ref{fig:umap_d2o_A}, 8.2\% of events): D$_2$O$^{2+}$ ($^{1}A_{2}$) dissociating to D$^{+}$ + D$^{+}$ + O($^{1}$D) 
\begin{itemize}
\item A singlet dication state with a peak KER of $\sim$ 7.6 eV and a $\beta$ of $\sim$ 126$^{\circ}$.
\end{itemize}

\textbf{Cluster 5} (orange in Figs~\ref{fig:umap_d2o_A}~-~\ref{fig:umap_d2o_D}, 10.7\% of events): D$_2$O$^{2+}$ (2$^{1}A_{1}$) dissociating to D$^{+}$ + D$^{+}$ + O($^{1}$D) 
\begin{itemize}
\item A singlet dication state characterized by a high $\beta$ (protons nearly opposite directions) and a peak KER of $\sim$ 7.7 eV. 
\end{itemize}

\textbf{Cluster 6} (light-blue in Figs~\ref{fig:umap_d2o_A}~-~\ref{fig:umap_d2o_D}, 18.2\% of events): D$_2$O$^{2+}$ ($^{1}B_{2}$) dissociating to D$^{+}$ + D$^{+}$ + O($^{1}$D) 
\begin{itemize}
\item A singlet dication state with a peak KER of $\sim$ 9.8 eV and a $\beta$ of $\sim$ 143$^{\circ}$. It is a strong contributor to the O($^{1}$D) asymptote.
\end{itemize}

\textbf{Cluster 7} (pink in Fig.~\ref{fig:umap_d2o_A}, 19\% of events): D$_2$O$^{2+}$ ($^{3}A_{2}$) dissociating to D$^{+}$ + D$^{+}$ + O($^{3}$P) 
\begin{itemize}
\item A triplet dication state with a peak KER of $\sim$ 8 eV and a $\beta$ of $\sim$ 121$^{\circ}$.
\end{itemize}

\textbf{Cluster 8} (light-green in Figs~\ref{fig:umap_d2o_A}~-~\ref{fig:umap_d2o_D}, 33.7\% of events): D$_2$O$^{2+}$ ($^{3}B_{2}$) dissociating to D$^{+}$ + D$^{+}$ + O($^{3}$P) 
\begin{itemize}
\item A triplet dication state with a peak KER of $\sim$ 9.7 eV and a $\beta$ of $\sim$ 139$^{\circ}$. It is a strong contributor to the O($^{3}$P) asymptote.
\end{itemize}

\subsubsection{Quality Metrics}

SCULPT's adaptive confidence scoring provided the following assessment for the initial UMAP analysis in Fig.~\ref{fig:umap_d2o_A}. These values can vary slightly from run to run due to the random sampling approach.

\begin{itemize}
\item Overall confidence: 0.71 (High reliability)
\item Silhouette score: 0.1324
\item Hopkins statistic: 0.9769
\item Stability: 0.9996
\item Physics consistency: 0.3184
\item Calinski Harabasz: 2338.4791
\item Davies Bouldin: 0.7306
\end{itemize}

\begin{figure}[htbp]
\centering
\includegraphics[width=\columnwidth]{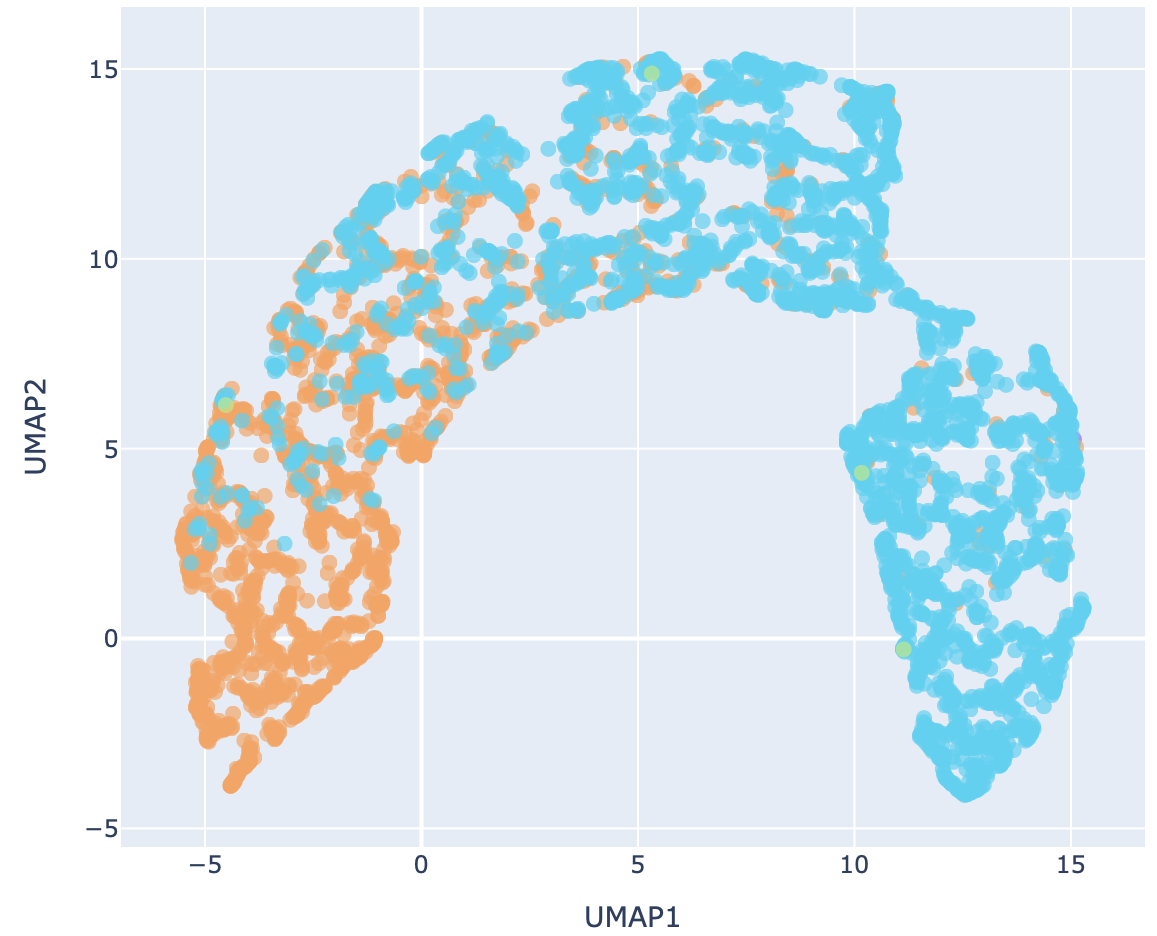}
\caption{UMAP projection of D$_2$O double ionization data showing poor clustering resulting from selecting only KER, EESum and Total Energy as UMAP features. The calculated confidence score is 0.14. Colors represent different dication states and are for visualization purposes only.}
\label{fig:wrong_umap_d2o_D}
\end{figure}

As clusters containing multiple quantum states are progressively separated through iterative analysis, the reliability score increases. For example, the confidence score for the UMAP embedding in Fig.~\ref{fig:umap_d2o_C} is 0.70, and when it is further sub-clustered, as shown in Fig.~\ref{fig:umap_d2o_D}, the confidence score becomes 0.79. In contrast, selecting only KER, EESum and Total Energy as UMAP features results in poor separation, as shown in Fig.~\ref{fig:wrong_umap_d2o_D}, and a poor score of 0.14. Hence, the confidence score, combined with visual inspection of cluster separation, guides the iterative selection of data subsets and appropriate features for subsequent clustering refinement.  

\textcolor{black}{It is important to note that individual metrics may exhibit low absolute values while the overall clustering remains physically meaningful. This is expected to be common in COLTRIMS data where quantum states exhibit inherent overlap in momentum space. For example, the silhouette score of 0.13 observed in Fig. ~\ref{fig:umap_d2o_A} reflects the fact that several dication states share similar kinematic signatures (e.g., the $^{3}B_{1}$ and $^{1}B_{1}$ states both have peak KER $\sim$ 4.3 eV), resulting in genuinely overlapping clusters in the UMAP projection. The confidence score accounts for this by weighting multiple complementary metrics—notably the Hopkins statistic (0.98 in Fig. ~\ref{fig:umap_d2o_A}), which indicates strong underlying clustering tendency despite the modest silhouette value. The iterative refinement workflow demonstrated in Section~\ref{section:casestudy} shows how progressively isolating clusters leads to improved confidence scores (0.70→0.79) as overlapping states are separated.}

\section{Conclusions}

SCULPT represents a significant advance in the analysis of multi-particle coincidence data from momentum spectroscopy experiments. By combining modern machine learning techniques with physics-based parameters for dimensionality reduction, the software enables researchers to extract detailed information from highly-dimensional data more efficiently than previously established analytical tools. We expect that SCULPT could also enable the identification and isolation of quantum states and dynamical mechanisms that could otherwise be left hidden in congested momentum spectra. The adaptive confidence scoring system provides quantitative reliability assessments, addressing a critical need for objective quality metrics in exploratory data analysis.

The case study of D$_2$O double ionization demonstrates SCULPT's ability to identify subtle fragmentation channels, with several being diffuse or diluted by competing processes, and discover non-intuitive feature combinations that enhance cluster separation. 
The web-based implementation and modular architecture make the software accessible to the broader community while enabling customization for specific experimental requirements.  

\section{Future Directions}

We have not utilized the entire spectrum of capabilities and options of SCULPT in our analysis just yet. \textcolor{black}{Near f}uture studies on different datasets will explore other capabilities of SCULPT such as advanced data filtering and the implementation of deep autoencoders. As momentum spectroscopy techniques continue to be further developed, producing ever-larger and more complex datasets, tools like SCULPT will become essential for extracting physical insights.

\textcolor{black}{The next development step will incorporate digital-twin and agent-based capabilities. A digital-twin simulation module, based on a classical Newtonian dissociation model, will predict experimental outcomes from user input and/or outputs of the Data Analysis \& Featurization module. This module will guide the clustering process, and together they will form an interactive machine-learning loop that supports the validation and interpretation of measured results while identifying relevant features and underlying physical processes.}

\textcolor{black}{In particular, the agent will be employed to isolate and characterize dissociation dynamics in the concerted fragmentation pathway of water that remained inaccessible using conventional, by-hand experimental analysis. Only detailed \textit{ab initio} theoretical investigations revealed that three of the eight water dication states exhibit non-standard dissociation behavior associated with a breakdown of the axial-recoil-approximation. This behavior arises from a so-called “slingshot mechanism,” which inverts the kinematics of the reaction products~\cite{streeter_dissociation_2018}. Notably, this effect could not be identified through traditional manual analysis of the experimental data alone.}

\textcolor{black}{Preliminary investigations using SCULPT indicate that the three clusters corresponding to these dication states exhibit varying degrees of overlap or separation in UMAP representations, depending on the choice of correlated features. The agent will be tasked with autonomously analyzing these UMAP patterns in conjunction with the digital-twin to identify shared characteristics among the clusters in feature space. This approach provides a data-driven pathway to detect events that violate the axial-recoil-approximation without relying on \textit{ab initio} calculations, and it enables the systematic identification of non-axial fragmentation dynamics in complex molecular systems that are beyond the reach of such sophisticated theoretical methods.}

\textcolor{black}{As such}, the SCULPT platform exemplifies a transformative approach to analyzing high-dimensional tabulated data, with broad implications across both fundamental science and industry. \textcolor{black}{We believe that} this work lays the foundation for AI-driven discovery in multi-particle quantum dynamics, emphasizing real-time, physics-informed analysis of complex experimental datasets. \textcolor{black}{In the long term, i}ts modular, web-based architecture and adaptive clustering tools \textcolor{black}{will} not only accelerate insight in COLTRIMS experiments but also \textcolor{black}{potentially} offer scalable solutions for applications and industries where uncovering rare events and nonlinear correlations in large tabulated multi-parameter datasets are critical, such as quantum information science,\cite{sennary_attosecond_2025} pharmaceuticals,\cite{berlin_adverse_2008,feng_comparison_2020,sadybekov_computational_2023} medicine,\cite{huang_application_2023} aerospace. \cite{srivastava_enabling_2006,janakiraman_anomaly_2016}
SCULPT can bridge experimental physics with data-centric innovation, positioning itself as a model for cross-sector impact. We have made SCULPT available on GitHub to encourage community contributions and to ensure that the software can evolve with the field's needs.

\begin{acknowledgments}
This work was supported by the Laboratory Directed Research and Development (LDRD) program at Lawrence Berkeley National Laboratory. We acknowledge the computational resources provided by the Advanced Light Source and the National Energy Research Scientific Computing Center (NERSC), which both are a DOE Office of Science User Facility under contract no. DE-AC02-05CH11231. In particular we acknowledge NERSC award ERCAP-0031498. We also thank the ALS Photon Science Computing Group for their valuable feedback.
\end{acknowledgments}

\section*{Data Availability}

The data that support the findings of this study are available from the corresponding author upon reasonable request. Example datasets and tutorials are available at \textcolor{black}{\url{https://github.com/AMOS-experiment/CoInML/}}.

\bibliography{SCULPT}

@article{dorner_cold_2000,
	title = {Cold {Target} {Recoil} {Ion} {Momentum} {Spectroscopy}: a ‘momentum microscope’ to view atomic collision dynamics},
	volume = {330},
	copyright = {https://www.elsevier.com/tdm/userlicense/1.0/},
	issn = {03701573},
	shorttitle = {Cold {Target} {Recoil} {Ion} {Momentum} {Spectroscopy}},
	url = {https://linkinghub.elsevier.com/retrieve/pii/S037015739900109X},
	doi = {10.1016/S0370-1573(99)00109-X},
	nolanguage = {en},
	number = {2-3},
	urldate = {2026-01-22},
	journal = {Physics Reports},
	author = {Dörner, R. and Mergel, V. and Jagutzki, O. and Spielberger, L. and Ullrich, J. and Moshammer, R. and Schmidt-Böcking, H.},
	month = jun,
	year = {2000},
	pages = {95--192},
}

@article{ullrich_recoil-ion_2003,
	title = {Recoil-ion and electron momentum spectroscopy: reaction-microscopes},
	volume = {66},
	issn = {0034-4885, 1361-6633},
	shorttitle = {Recoil-ion and electron momentum spectroscopy},
	url = {https://iopscience.iop.org/article/10.1088/0034-4885/66/9/203},
	doi = {10.1088/0034-4885/66/9/203},
	number = {9},
	urldate = {2026-01-22},
	journal = {Reports on Progress in Physics},
	author = {Ullrich, J and Moshammer, R and Dorn, A and D Rner, R and Schmidt, L Ph H and Schmidt-B Cking, H},
	month = sep,
	year = {2003},
	pages = {1463--1545},
}

@article{jahnke_photoelectron_2007,
	title = {Photoelectron and {ICD} electron angular distributions from fixed-in-space neon dimers},
	volume = {40},
	issn = {0953-4075, 1361-6455},
	url = {https://iopscience.iop.org/article/10.1088/0953-4075/40/13/006},
	doi = {10.1088/0953-4075/40/13/006},
	number = {13},
	urldate = {2026-01-22},
	journal = {Journal of Physics B: Atomic, Molecular and Optical Physics},
	author = {Jahnke, T and Czasch, A and Schöffler, M and Schössler, S and Käsz, M and Titze, J and Kreidi, K and Grisenti, R E and Staudte, A and Jagutzki, O and Schmidt, L Ph H and Semenov, S K and Cherepkov, N A and Schmidt-Böcking, H and Dörner, R},
	month = jul,
	year = {2007},
	pages = {2597--2606},
}

@article{schoffler_ultrafast_2008,
	title = {Ultrafast {Probing} of {Core} {Hole} {Localization} in {N}$_{\textrm{2}}$},
	volume = {320},
	issn = {0036-8075, 1095-9203},
	url = {https://www.science.org/doi/10.1126/science.1154989},
	doi = {10.1126/science.1154989},
	nolanguage = {en},
	number = {5878},
	urldate = {2026-01-23},
	journal = {Science},
	author = {Schöffler, M. S. and Titze, J. and Petridis, N. and Jahnke, T. and Cole, K. and Schmidt, L. Ph. H. and Czasch, A. and Akoury, D. and Jagutzki, O. and Williams, J. B. and Cherepkov, N. A. and Semenov, S. K. and McCurdy, C. W. and Rescigno, T. N. and Cocke, C. L. and Osipov, T. and Lee, S. and Prior, M. H. and Belkacem, A. and Landers, A. L. and Schmidt-Böcking, H. and Weber, Th. and Dörner, R.},
	month = may,
	year = {2008},
	pages = {920--923},
}

@article{weber_complete_2004,
	title = {Complete photo-fragmentation of the deuterium molecule},
	volume = {431},
	copyright = {http://www.springer.com/tdm},
	issn = {0028-0836, 1476-4687},
	url = {https://www.nature.com/articles/nature02839},
	doi = {10.1038/nature02839},
	nolanguage = {en},
	number = {7007},
	urldate = {2026-01-23},
	journal = {Nature},
	author = {Weber, T. and Czasch, A. O. and Jagutzki, O. and Müller, A. K. and Mergel, V. and Kheifets, A. and Rotenberg, E. and Meigs, G. and Prior, M. H. and Daveau, S. and Landers, A. and Cocke, C. L. and Osipov, T. and Díez Muiño, R. and Schmidt-Böcking, H. and Dörner, R.},
	month = sep,
	year = {2004},
	pages = {437--440},
}

@article{fehre_absolute_2018,
	title = {Absolute ion detection efficiencies of microchannel plates and funnel microchannel plates for multi-coincidence detection},
	volume = {89},
	issn = {0034-6748, 1089-7623},
	url = {https://pubs.aip.org/rsi/article/89/4/045112/362091/Absolute-ion-detection-efficiencies-of},
	doi = {10.1063/1.5022564},
	nolanguage = {en},
	number = {4},
	urldate = {2026-01-23},
	journal = {Review of Scientific Instruments},
	author = {Fehre, K. and Trojanowskaja, D. and Gatzke, J. and Kunitski, M. and Trinter, F. and Zeller, S. and Schmidt, L. Ph. H. and Stohner, J. and Berger, R. and Czasch, A. and Jagutzki, O. and Jahnke, T. and Dörner, R. and Schöffler, M. S.},
	month = apr,
	year = {2018},
	pages = {045112},
}

@article{demekhin_exploring_2011,
	title = {Exploring {Interatomic} {Coulombic} {Decay} by {Free} {Electron} {Lasers}},
	volume = {107},
	copyright = {http://link.aps.org/licenses/aps-default-license},
	issn = {0031-9007, 1079-7114},
	url = {https://link.aps.org/doi/10.1103/PhysRevLett.107.273002},
	doi = {10.1103/PhysRevLett.107.273002},
	nolanguage = {en},
	number = {27},
	urldate = {2026-01-23},
	journal = {Physical Review Letters},
	author = {Demekhin, Philipp V. and Stoychev, Spas D. and Kuleff, Alexander I. and Cederbaum, Lorenz S.},
	month = dec,
	year = {2011},
	pages = {273002},
}

@article{czasch_partial_2005,
	title = {Partial {Photoionization} {Cross} {Sections} and {Angular} {Distributions} for {Double} {Excitation} of {Helium} up to the {N} = 13 {Threshold}},
	volume = {95},
	copyright = {http://link.aps.org/licenses/aps-default-license},
	issn = {0031-9007, 1079-7114},
	url = {https://link.aps.org/doi/10.1103/PhysRevLett.95.243003},
	doi = {10.1103/PhysRevLett.95.243003},
	nolanguage = {en},
	number = {24},
	urldate = {2026-01-23},
	journal = {Physical Review Letters},
	author = {Czasch, A. and Schöffler, M. and Hattass, M. and Schössler, S. and Jahnke, T. and Weber, Th. and Staudte, A. and Titze, J. and Wimmer, C. and Kammer, S. and Weckenbrock, M. and Voss, S. and Grisenti, R. E. and Jagutzki, O. and Schmidt, L. Ph. H. and Schmidt-Böcking, H. and Dörner, R. and Rost, J. M. and Schneider, T. and Liu, Chien-Nan and Bray, I. and Kheifets, A. S. and Bartschat, K.},
	month = dec,
	year = {2005},
	pages = {243003},
}

@article{schmidt_spatial_2012,
	title = {Spatial {Imaging} of the {H} 2 + {Vibrational} {Wave} {Function} at the {Quantum} {Limit}},
	volume = {108},
	copyright = {http://link.aps.org/licenses/aps-default-license},
	issn = {0031-9007, 1079-7114},
	url = {https://link.aps.org/doi/10.1103/PhysRevLett.108.073202},
	doi = {10.1103/PhysRevLett.108.073202},
	nolanguage = {en},
	number = {7},
	urldate = {2026-01-23},
	journal = {Physical Review Letters},
	author = {Schmidt, L. Ph. H. and Jahnke, T. and Czasch, A. and Schöffler, M. and Schmidt-Böcking, H. and Dörner, R.},
	month = feb,
	year = {2012},
	pages = {073202},
}

@article{lecun_deep_2015,
	title = {Deep learning},
	volume = {521},
	issn = {0028-0836, 1476-4687},
	url = {https://www.nature.com/articles/nature14539},
	doi = {10.1038/nature14539},
	nolanguage = {en},
	number = {7553},
	urldate = {2026-01-23},
	journal = {Nature},
	author = {LeCun, Yann and Bengio, Yoshua and Hinton, Geoffrey},
	month = may,
	year = {2015},
	pages = {436--444},
}

@article{wang_scientific_2023,
	title = {Scientific discovery in the age of artificial intelligence},
	volume = {620},
	issn = {0028-0836, 1476-4687},
	url = {https://www.nature.com/articles/s41586-023-06221-2},
	doi = {10.1038/s41586-023-06221-2},
	nolanguage = {en},
	number = {7972},
	urldate = {2026-01-23},
	journal = {Nature},
	author = {Wang, Hanchen and Fu, Tianfan and Du, Yuanqi and Gao, Wenhao and Huang, Kexin and Liu, Ziming and Chandak, Payal and Liu, Shengchao and Van Katwyk, Peter and Deac, Andreea and Anandkumar, Anima and Bergen, Karianne and Gomes, Carla P. and Ho, Shirley and Kohli, Pushmeet and Lasenby, Joan and Leskovec, Jure and Liu, Tie-Yan and Manrai, Arjun and Marks, Debora and Ramsundar, Bharath and Song, Le and Sun, Jimeng and Tang, Jian and Veličković, Petar and Welling, Max and Zhang, Linfeng and Coley, Connor W. and Bengio, Yoshua and Zitnik, Marinka},
	month = aug,
	year = {2023},
	pages = {47--60},
}

@article{chen_machine_2021,
	title = {Machine learning on neutron and x-ray scattering and spectroscopies},
	volume = {2},
	issn = {2688-4070},
	url = {https://pubs.aip.org/cpr/article/2/3/031301/138019/Machine-learning-on-neutron-and-x-ray-scattering},
	doi = {10.1063/5.0049111},
	nolanguage = {en},
	number = {3},
	urldate = {2026-01-23},
	journal = {Chemical Physics Reviews},
	author = {Chen, Zhantao and Andrejevic, Nina and Drucker, Nathan C. and Nguyen, Thanh and Xian, R. Patrick and Smidt, Tess and Wang, Yao and Ernstorfer, Ralph and Tennant, D. Alan and Chan, Maria and Li, Mingda},
	month = sep,
	year = {2021},
	pages = {031301},
}

@article{hu_deep_2025,
	title = {Deep learning for ultrafast {X}-ray scattering and imaging with intense {X}-ray {FEL} pulses},
	volume = {14},
	issn = {2192-8584},
	url = {https://www.frontiersin.org/articles/10.3389/aot.2025.1546386/full},
	doi = {10.3389/aot.2025.1546386},
	urldate = {2026-01-23},
	journal = {Advanced Optical Technologies},
	author = {Hu, Menglu and Fan, Jiadong and Tong, Yajun and Sun, Zhibin and Jiang, Huaidong},
	month = mar,
	year = {2025},
	pages = {1546386},
}

@article{daoud_novel_2023,
	title = {Novel applications of generative adversarial networks ({GANs}) in the analysis of ultrafast electron diffraction ({UED}) images},
	volume = {159},
	issn = {0021-9606, 1089-7690},
	url = {https://pubs.aip.org/jcp/article/159/4/044107/2904225/Novel-applications-of-generative-adversarial},
	doi = {10.1063/5.0154871},
	nolanguage = {en},
	number = {4},
	urldate = {2026-01-23},
	journal = {The Journal of Chemical Physics},
	author = {Daoud, Hazem and Sirohi, Dhruv and Mjeku, Endri and Feng, John and Oghbaey, Saeed and Miller, R. J. Dwayne},
	month = jul,
	year = {2023},
	pages = {044107},
}

@article{zhao_visualization_2023,
	title = {Visualization of small-angle {X}-ray scattering datasets and processing-structure mapping of isotactic polypropylene films by machine learning},
	volume = {228},
	issn = {02641275},
	url = {https://linkinghub.elsevier.com/retrieve/pii/S0264127523002435},
	doi = {10.1016/j.matdes.2023.111828},
	nolanguage = {en},
	urldate = {2026-01-23},
	journal = {Materials \& Design},
	author = {Zhao, Chenhao and Yu, Wancheng and Li, Liangbin},
	month = apr,
	year = {2023},
	pages = {111828},
}

@article{he_model_2020,
	title = {Model {Reconstruction} from {Small}-{Angle} {X}-{Ray} {Scattering} {Data} {Using} {Deep} {Learning} {Methods}},
	volume = {23},
	issn = {25890042},
	url = {https://linkinghub.elsevier.com/retrieve/pii/S2589004220300900},
	doi = {10.1016/j.isci.2020.100906},
	nolanguage = {en},
	number = {3},
	urldate = {2026-01-23},
	journal = {iScience},
	author = {He, Hao and Liu, Can and Liu, Haiguang},
	month = mar,
	year = {2020},
	pages = {100906},
}

@article{chavez_machine-learning-driven_2025,
	title = {A machine-learning-driven data labeling pipeline for scientific analysis in \textit{{MLExchange}}},
	volume = {58},
	issn = {1600-5767},
	url = {https://journals.iucr.org/paper?S1600576725002328},
	doi = {10.1107/S1600576725002328},
	number = {3},
	urldate = {2026-01-23},
	journal = {Journal of Applied Crystallography},
	author = {Chavez, Tanny and Zhao, Zhuowen and Jiang, Runbo and Koepp, Wiebke and McReynolds, Dylan and Zwart, Petrus H. and Allan, Daniel B. and Gann, Eliot H. and Schwarz, Nicholas and Ushizima, Daniela and Barnard, Edward S. and Mehta, Apurva and Sankaranarayanan, Subramanian and Hexemer, Alexander},
	month = jun,
	year = {2025},
	pages = {731--745},
}

@article{abdelmoula_peak_2021,
	title = {Peak learning of mass spectrometry imaging data using artificial neural networks},
	volume = {12},
	issn = {2041-1723},
	url = {https://www.nature.com/articles/s41467-021-25744-8},
	doi = {10.1038/s41467-021-25744-8},
	nolanguage = {en},
	number = {1},
	urldate = {2026-01-23},
	journal = {Nature Communications},
	author = {Abdelmoula, Walid M. and Lopez, Begona Gimenez-Cassina and Randall, Elizabeth C. and Kapur, Tina and Sarkaria, Jann N. and White, Forest M. and Agar, Jeffrey N. and Wells, William M. and Agar, Nathalie Y. R.},
	month = sep,
	year = {2021},
	pages = {5544},
}

@article{bitto_enhancing_2024,
	title = {Enhancing mass spectrometry imaging accessibility using convolutional autoencoders for deriving hypoxia-associated peptides from tumors},
	volume = {10},
	issn = {2056-7189},
	url = {https://www.nature.com/articles/s41540-024-00385-x},
	doi = {10.1038/s41540-024-00385-x},
	nolanguage = {en},
	number = {1},
	urldate = {2026-01-23},
	journal = {npj Systems Biology and Applications},
	author = {Bitto, Verena and Hönscheid, Pia and Besso, María José and Sperling, Christian and Kurth, Ina and Baumann, Michael and Brors, Benedikt},
	month = may,
	year = {2024},
	pages = {57},
}

@article{becht_dimensionality_2019,
	title = {Dimensionality reduction for visualizing single-cell data using {UMAP}},
	volume = {37},
	issn = {1087-0156, 1546-1696},
	url = {https://www.nature.com/articles/nbt.4314},
	doi = {10.1038/nbt.4314},
	nolanguage = {en},
	number = {1},
	urldate = {2026-01-23},
	journal = {Nature Biotechnology},
	author = {Becht, Etienne and McInnes, Leland and Healy, John and Dutertre, Charles-Antoine and Kwok, Immanuel W H and Ng, Lai Guan and Ginhoux, Florent and Newell, Evan W},
	month = jan,
	year = {2019},
	pages = {38--44},
}

@article{szubert_structure-preserving_2019,
	title = {Structure-preserving visualisation of high dimensional single-cell datasets},
	volume = {9},
	issn = {2045-2322},
	url = {https://www.nature.com/articles/s41598-019-45301-0},
	doi = {10.1038/s41598-019-45301-0},
	nolanguage = {en},
	number = {1},
	urldate = {2026-01-23},
	journal = {Scientific Reports},
	author = {Szubert, Benjamin and Cole, Jennifer E. and Monaco, Claudia and Drozdov, Ignat},
	month = jun,
	year = {2019},
	pages = {8914},
}

@article{trozzi_umap_2021,
	title = {{UMAP} as a {Dimensionality} {Reduction} {Tool} for {Molecular} {Dynamics} {Simulations} of {Biomacromolecules}: {A} {Comparison} {Study}},
	volume = {125},
	copyright = {https://doi.org/10.15223/policy-029},
	issn = {1520-6106, 1520-5207},
	shorttitle = {{UMAP} as a {Dimensionality} {Reduction} {Tool} for {Molecular} {Dynamics} {Simulations} of {Biomacromolecules}},
	url = {https://pubs.acs.org/doi/10.1021/acs.jpcb.1c02081},
	doi = {10.1021/acs.jpcb.1c02081},
	nolanguage = {en},
	number = {19},
	urldate = {2026-01-23},
	journal = {The Journal of Physical Chemistry B},
	author = {Trozzi, Francesco and Wang, Xinlei and Tao, Peng},
	month = may,
	year = {2021},
	pages = {5022--5034},
}

@article{bourilkov_machine_2019,
	title = {Machine and deep learning applications in particle physics},
	volume = {34},
	issn = {0217-751X, 1793-656X},
	url = {https://www.worldscientific.com/doi/abs/10.1142/S0217751X19300199},
	doi = {10.1142/S0217751X19300199},
	nolanguage = {en},
	number = {35},
	urldate = {2026-01-23},
	journal = {International Journal of Modern Physics A},
	author = {Bourilkov, Dimitri},
	month = dec,
	year = {2019},
	pages = {1930019},
}

@article{guest_deep_2018,
	title = {Deep {Learning} and {Its} {Application} to {LHC} {Physics}},
	volume = {68},
	issn = {0163-8998, 1545-4134},
	url = {https://www.annualreviews.org/doi/10.1146/annurev-nucl-101917-021019},
	doi = {10.1146/annurev-nucl-101917-021019},
	nolanguage = {en},
	number = {1},
	urldate = {2026-01-23},
	journal = {Annual Review of Nuclear and Particle Science},
	author = {Guest, Dan and Cranmer, Kyle and Whiteson, Daniel},
	month = oct,
	year = {2018},
	pages = {161--181},
}

@article{venkatachalam_exploiting_2025,
	title = {Exploiting correlations in multi-coincidence {Coulomb} explosion patterns for differentiating molecular structures using machine learning},
	volume = {16},
	copyright = {2025 The Author(s)},
	issn = {2041-1723},
	url = {https://www.nature.com/articles/s41467-025-66369-5},
	doi = {10.1038/s41467-025-66369-5},
	language = {en},
	number = {1},
	urldate = {2026-02-18},
	journal = {Nature Communications},
	publisher = {Nature Publishing Group},
	author = {Venkatachalam, Anbu Selvam and Greenman, Loren and Stallbaumer, Joshua and Rudenko, Artem and Rolles, Daniel and Lam, Huynh Van Sa},
	month = dec,
	year = {2025},
	pages = {11366},
}

@article{li_generative_2026,
	title = {Generative modeling enables molecular structure retrieval from {Coulomb} explosion imaging},
	volume = {17},
	copyright = {2026 The Author(s)},
	issn = {2041-1723},
	url = {https://www.nature.com/articles/s41467-026-70160-5},
	doi = {10.1038/s41467-026-70160-5},
	language = {en},
	number = {1},
	urldate = {2026-05-14},
	journal = {Nature Communications},
	publisher = {Nature Publishing Group},
	author = {Li, Xiang and Jahnke, Till and Boll, Rebecca and Han, Jiaqi and Xu, Minkai and Meyer, Michael and Piancastelli, Maria Novella and Rolles, Daniel and Rudenko, Artem and Trinter, Florian and Wolf, Thomas J. A. and Thayer, Jana B. and Cryan, James P. and Ermon, Stefano and Ho, Phay J.},
	month = mar,
	year = {2026},
	pages = {3430},
}

@misc{mcinnes_umap_2020,
	title = {{UMAP}: {Uniform} {Manifold} {Approximation} and {Projection} for {Dimension} {Reduction}},
	shorttitle = {{UMAP}},
	url = {http://arxiv.org/abs/1802.03426},
	doi = {10.48550/arXiv.1802.03426},
	urldate = {2026-02-18},
	publisher = {arXiv},
	author = {McInnes, Leland and Healy, John and Melville, James},
	month = sep,
	year = {2020},
	note = {arXiv:1802.03426 [stat]},
}

@article{fortin_deap_2012,
	title = {{DEAP}: evolutionary algorithms made easy},
	volume = {13},
	issn = {1532-4435},
	shorttitle = {{DEAP}},
	url = {https://dl.acm.org/doi/10.5555/2503308.2503311},
	number = {1},
	urldate = {2026-01-23},
	journal = {J. Mach. Learn. Res.},
	author = {Fortin, Félix-Antoine and De Rainville, François-Michel and Gardner, Marc-André Gardner and Parizeau, Marc and Gagné, Christian},
	month = jul,
	year = {2012},
	pages = {2171--2175},
}

@misc{brehmer_hierarchical_2020,
	title = {Hierarchical clustering in particle physics through reinforcement learning},
	url = {http://arxiv.org/abs/2011.08191},
	doi = {10.48550/arXiv.2011.08191},
	urldate = {2026-02-18},
	publisher = {arXiv},
	author = {Brehmer, Johann and Macaluso, Sebastian and Pappadopulo, Duccio and Cranmer, Kyle},
	month = dec,
	year = {2020},
	note = {arXiv:2011.08191 [cs]},
}

@inproceedings{cherrier_consistent_2019,
	title = {Consistent {Feature} {Construction} with {Constrained} {Genetic} {Programming} for {Experimental} {Physics}},
	url = {https://ieeexplore.ieee.org/document/8789937},
	doi = {10.1109/CEC.2019.8789937},
	urldate = {2026-01-23},
	booktitle = {2019 {IEEE} {Congress} on {Evolutionary} {Computation} ({CEC})},
	author = {Cherrier, Noëlie and Poli, Jean-Philippe and Defurne, Maxime and Sabatié, Franck},
	month = jun,
	year = {2019},
	pages = {1650--1658},
}

@article{lacava_learning_2020,
	title = {Learning feature spaces for regression with genetic programming},
	volume = {21},
	issn = {1389-2576, 1573-7632},
	url = {http://link.springer.com/10.1007/s10710-020-09383-4},
	doi = {10.1007/s10710-020-09383-4},
	nolanguage = {en},
	number = {3},
	urldate = {2026-01-23},
	journal = {Genetic Programming and Evolvable Machines},
	author = {La Cava, William and Moore, Jason H.},
	month = sep,
	year = {2020},
	pages = {433--467},
}

@article{makke_interpretable_2024,
	title = {Interpretable scientific discovery with symbolic regression: a review},
	volume = {57},
	issn = {0269-2821, 1573-7462},
	shorttitle = {Interpretable scientific discovery with symbolic regression},
	url = {https://link.springer.com/10.1007/s10462-023-10622-0},
	doi = {10.1007/s10462-023-10622-0},
	nolanguage = {en},
	number = {1},
	urldate = {2026-01-23},
	journal = {Artificial Intelligence Review},
	author = {Makke, Nour and Chawla, Sanjay},
	month = jan,
	year = {2024},
	pages = {2},
}

@article{reedy_dissociation_2018,
	title = {Dissociation dynamics of the water dication following one-photon double ionization. {II}. {Experiment}},
	volume = {98},
	issn = {2469-9926, 2469-9934},
	url = {https://link.aps.org/doi/10.1103/PhysRevA.98.053430},
	doi = {10.1103/PhysRevA.98.053430},
	nolanguage = {en},
	number = {5},
	urldate = {2026-01-23},
	journal = {Physical Review A},
	author = {Reedy, D. and Williams, J. B. and Gaire, B. and Gatton, A. and Weller, M. and Menssen, A. and Bauer, T. and Henrichs, K. and Burzynski, Ph. and Berry, B. and Streeter, Z. L. and Sartor, J. and Ben-Itzhak, I. and Jahnke, T. and Dörner, R. and Weber, Th. and Landers, A. L.},
	month = nov,
	year = {2018},
	pages = {053430},
}

@article{rousseeuw_silhouettes_1987,
	title = {Silhouettes: {A} graphical aid to the interpretation and validation of cluster analysis},
	volume = {20},
	copyright = {https://www.elsevier.com/tdm/userlicense/1.0/},
	issn = {03770427},
	shorttitle = {Silhouettes},
	url = {https://linkinghub.elsevier.com/retrieve/pii/0377042787901257},
	doi = {10.1016/0377-0427(87)90125-7},
	nolanguage = {en},
	urldate = {2026-01-23},
	journal = {Journal of Computational and Applied Mathematics},
	author = {Rousseeuw, Peter J.},
	month = nov,
	year = {1987},
	pages = {53--65},
}

@article{hopkins_new_1954,
	title = {A {New} {Method} for determining the {Type} of {Distribution} of {Plant} {Individuals}},
	volume = {18},
	issn = {1095-8290, 0305-7364},
	url = {https://academic.oup.com/aob/article/277917/A},
	doi = {10.1093/oxfordjournals.aob.a083391},
	nolanguage = {en},
	number = {2},
	urldate = {2026-01-23},
	journal = {Annals of Botany},
	author = {Hopkins, Brian and Skellam, J. G.},
	month = apr,
	year = {1954},
	pages = {213--227},
}

@article{lawson_new_1990,
	title = {New index for clustering tendency and its application to chemical problems},
	volume = {30},
	issn = {0095-2338, 1520-5142},
	url = {https://pubs.acs.org/doi/abs/10.1021/ci00065a010},
	doi = {10.1021/ci00065a010},
	nolanguage = {en},
	number = {1},
	urldate = {2026-01-23},
	journal = {Journal of Chemical Information and Computer Sciences},
	author = {Lawson, Richard G. and Jurs, Peter C.},
	month = feb,
	year = {1990},
	pages = {36--41},
}

@article{hubert_comparing_1985,
	title = {Comparing partitions},
	volume = {2},
	copyright = {http://www.springer.com/tdm},
	issn = {0176-4268, 1432-1343},
	url = {http://link.springer.com/10.1007/BF01908075},
	doi = {10.1007/BF01908075},
	nolanguage = {en},
	number = {1},
	urldate = {2026-01-23},
	journal = {Journal of Classification},
	author = {Hubert, Lawrence and Arabie, Phipps},
	month = dec,
	year = {1985},
	pages = {193--218},
}

@article{calinski_dendrite_1974,
	title = {A dendrite method for cluster analysis},
	volume = {3},
	issn = {0361-0926},
	url = {http://www.tandfonline.com/doi/abs/10.1080/03610927408827101},
	doi = {10.1080/03610927408827101},
	nolanguage = {en},
	number = {1},
	urldate = {2026-01-23},
	journal = {Communications in Statistics - Theory and Methods},
	author = {Calinski, T. and Harabasz, J.},
	year = {1974},
	pages = {1--27},
}

@article{davies_cluster_1979,
	title = {A {Cluster} {Separation} {Measure}},
	volume = {PAMI-1},
	copyright = {https://ieeexplore.ieee.org/Xplorehelp/downloads/license-information/IEEE.html},
	issn = {0162-8828, 2160-9292},
	url = {http://ieeexplore.ieee.org/document/4766909/},
	doi = {10.1109/TPAMI.1979.4766909},
	number = {2},
	urldate = {2026-01-23},
	journal = {IEEE Transactions on Pattern Analysis and Machine Intelligence},
	author = {Davies, David L. and Bouldin, Donald W.},
	month = apr,
	year = {1979},
	pages = {224--227},
}

@article{arbelaitz_extensive_2013,
	title = {An extensive comparative study of cluster validity indices},
	volume = {46},
	copyright = {https://www.elsevier.com/tdm/userlicense/1.0/},
	issn = {00313203},
	url = {https://linkinghub.elsevier.com/retrieve/pii/S003132031200338X},
	doi = {10.1016/j.patcog.2012.07.021},
	nolanguage = {en},
	number = {1},
	urldate = {2026-01-23},
	journal = {Pattern Recognition},
	author = {Arbelaitz, Olatz and Gurrutxaga, Ibai and Muguerza, Javier and Pérez, Jesús M. and Perona, Iñigo},
	month = jan,
	year = {2013},
	pages = {243--256},
}

@inproceedings{ester_density-based_1996,
	address = {Portland, Oregon},
	series = {{KDD}'96},
	title = {A density-based algorithm for discovering clusters in large spatial databases with noise},
	urldate = {2026-01-23},
	booktitle = {Proceedings of the {Second} {International} {Conference} on {Knowledge} {Discovery} and {Data} {Mining}},
	publisher = {AAAI Press},
	author = {Ester, Martin and Kriegel, Hans-Peter and Sander, Jörg and Xu, Xiaowei},
	month = aug,
	year = {1996},
	pages = {226--231},
}

@article{sennary_attosecond_2025,
	title = {Attosecond quantum uncertainty dynamics and ultrafast squeezed light for quantum communication},
	volume = {14},
	issn = {2047-7538},
	url = {https://www.nature.com/articles/s41377-025-02055-x},
	doi = {10.1038/s41377-025-02055-x},
	nolanguage = {en},
	number = {1},
	urldate = {2026-01-23},
	journal = {Light: Science \& Applications},
	author = {Sennary, Mohamed and Rivera-Dean, Javier and ElKabbash, Mohamed and Pervak, Vladimir and Lewenstein, Maciej and Hassan, Mohammed Th.},
	month = oct,
	year = {2025},
	pages = {350},
}

@article{berlin_adverse_2008,
	title = {Adverse {Event} {Detection} in {Drug} {Development}: {Recommendations} and {Obligations} {Beyond} {Phase} 3},
	volume = {98},
	issn = {0090-0036, 1541-0048},
	shorttitle = {Adverse {Event} {Detection} in {Drug} {Development}},
	url = {https://ajph.aphapublications.org/doi/full/10.2105/AJPH.2007.124537},
	doi = {10.2105/AJPH.2007.124537},
	nolanguage = {en},
	number = {8},
	urldate = {2026-01-23},
	journal = {American Journal of Public Health},
	author = {Berlin, Jesse A. and Glasser, Susan C. and Ellenberg, Susan S.},
	month = aug,
	year = {2008},
	pages = {1366--1371},
}

@article{feng_comparison_2020,
	title = {A comparison of residual diagnosis tools for diagnosing regression models for count data},
	volume = {20},
	issn = {1471-2288},
	url = {https://bmcmedresmethodol.biomedcentral.com/articles/10.1186/s12874-020-01055-2},
	doi = {10.1186/s12874-020-01055-2},
	nolanguage = {en},
	number = {1},
	urldate = {2026-01-23},
	journal = {BMC Medical Research Methodology},
	author = {Feng, Cindy and Li, Longhai and Sadeghpour, Alireza},
	month = dec,
	year = {2020},
	pages = {175},
}

@article{sadybekov_computational_2023,
	title = {Computational approaches streamlining drug discovery},
	volume = {616},
	issn = {0028-0836, 1476-4687},
	url = {https://www.nature.com/articles/s41586-023-05905-z},
	doi = {10.1038/s41586-023-05905-z},
	nolanguage = {en},
	number = {7958},
	urldate = {2026-01-23},
	journal = {Nature},
	author = {Sadybekov, Anastasiia V. and Katritch, Vsevolod},
	month = apr,
	year = {2023},
	pages = {673--685},
}

@article{huang_application_2023,
	title = {Application of machine learning in predicting survival outcomes involving real-world data: a scoping review},
	volume = {23},
	issn = {1471-2288},
	shorttitle = {Application of machine learning in predicting survival outcomes involving real-world data},
	url = {https://bmcmedresmethodol.biomedcentral.com/articles/10.1186/s12874-023-02078-1},
	doi = {10.1186/s12874-023-02078-1},
	nolanguage = {en},
	number = {1},
	urldate = {2026-01-23},
	journal = {BMC Medical Research Methodology},
	author = {Huang, Yinan and Li, Jieni and Li, Mai and Aparasu, Rajender R.},
	month = nov,
	year = {2023},
	pages = {268},
}

@inproceedings{srivastava_enabling_2006,
	address = {Big Sky, MT, USA},
	title = {Enabling the {Discovery} of {Recurring} {Anomalies} in {Aerospace} {Problem} {Reports} using {High}-{Dimensional} {Clustering} {Techniques}},
	isbn = {978-0-7803-9545-9},
	url = {http://ieeexplore.ieee.org/document/1656136/},
	doi = {10.1109/AERO.2006.1656136},
	urldate = {2026-01-23},
	booktitle = {2006 {IEEE} {Aerospace} {Conference}},
	publisher = {IEEE},
	author = {Srivastava, A.N. and Akella, R. and Diev, V. and Kumaresan, S.P. and McIntosh, D.M. and Pontikakis, E.D. and {Zuobing Xu} and {Yi Zhang}},
	year = {2006},
	pages = {1--17},
}

@inproceedings{janakiraman_anomaly_2016,
	address = {Vancouver, BC, Canada},
	title = {Anomaly detection in aviation data using extreme learning machines},
	isbn = {978-1-5090-0620-5},
	url = {http://ieeexplore.ieee.org/document/7727444/},
	doi = {10.1109/IJCNN.2016.7727444},
	urldate = {2026-01-23},
	booktitle = {2016 {International} {Joint} {Conference} on {Neural} {Networks} ({IJCNN})},
	publisher = {IEEE},
	author = {Janakiraman, Vijay Manikandan and Nielsen, David},
	month = jul,
	year = {2016},
	pages = {1993--2000},
}

@misc{plotly_technologies_inc_collaborative_2015,
	address = {Montreal, QC},
	title = {Collaborative {Data} {Science}},
	shorttitle = {Plotly},
	url = {https://plot.ly},
	author = {{Plotly Technologies Inc.}},
	year = {2015},
}

@article{streeter_dissociation_2018,
	title = {\textcolor{black}{Dissociation dynamics of the water dication following one-photon double ionization. {I}. {Theory}}},
	volume = {\textcolor{black}{98}},
	issn = {2469-9926, 2469-9934},
	url = {https://link.aps.org/doi/10.1103/PhysRevA.98.053429},
	doi = {\textcolor{black}{10.1103/PhysRevA.98.053429}},
	nolanguage = {en},
	number = {5},
	urldate = {2026-01-26},
	journal = {\textcolor{black}{Physical Review A}},
	author = {\textcolor{black}{Z. L. Streeter, F. Yip, R. L. Lucchese, B. Gervais, T. N. Rescigno, and C. W. Mc McCurdy}},
	month = nov,
	year = {\textcolor{black}{2018}}}

\end{document}